\documentclass[twocolumn,trackchanges,twocolappendix]{aastex701}

\usepackage{amsmath}
\usepackage{xspace}
\usepackage{enumitem}

\makeatletter 
  \patchcmd{\NAT@citex}
    {\@citea\NAT@hyper@{%
      \NAT@nmfmt{\NAT@nm}%
      \hyper@natlinkbreak{\NAT@aysep\NAT@spacechar}{\@citeb\@extra@b@citeb}%
      \NAT@date}}
    {\@citea\NAT@nmfmt{\NAT@nm}%
    \NAT@aysep\NAT@spacechar\NAT@hyper@{\NAT@date}}{}{}

  \patchcmd{\NAT@citex}
    {\@citea\NAT@hyper@{%
      \NAT@nmfmt{\NAT@nm}%
      \hyper@natlinkbreak{\NAT@spacechar\NAT@@open\if*#1*\else#1\NAT@spacechar\fi}%
        {\@citeb\@extra@b@citeb}%
      \NAT@date}}
    {\@citea\NAT@nmfmt{\NAT@nm}%
    \NAT@spacechar\NAT@@open\if*#1*\else#1\NAT@spacechar\fi\NAT@hyper@{\NAT@date}}
    {}{}
\makeatother

\newcommand{\xcrit}{x_{\mathrm{crit}}}
\newcommand{\Lya}{\mathrm{Ly}\alpha}
\DeclareUnicodeCharacter{03B1}{\ensuremath{\alpha}}

\newcommand{\pcm}{\, \mathrm{cm}^{-3}}
\newcommand{\Msolpc}{\, \mathrm{M}_{\sun} \, \mathrm{pc}^{-2}}

\newcommand{\Mcloud}{M_{\mathrm{cloud}}}

\newcommand{\Msun}{\mathrm{M}_{\sun}}

\newcommand{\ftrap}{f_{\mathrm{trap}}}

\newcommand{\agrav}{\mathbf{a_{\mathrm{grav}}}}
\newcommand{\aLya}{\mathbf{a_{\mathrm{Ly}\alpha}}}
\newcommand{\fEdd}{f_{\mathrm{Edd}}}

\newcommand\HI{{\text{H}\,\textsc{i}}\xspace} 
\newcommand\HII{{\text{H}\,\textsc{ii}}\xspace} 

\begin{document}

\title{Lyman-alpha Radiation Pressure in Dense Star Clusters: \\ Implications for Star Formation and Winds at Cosmic Dawn}

\author[0000-0002-0311-2206]{Shyam H. Menon}
\email{smenon@flatironinstitute.org}
\affiliation{Center for Computational Astrophysics, Flatiron Institute, 162 5th Avenue, New York, NY 10010, USA}
\affiliation{Department of Physics and Astronomy, Rutgers University, 136 Frelinghuysen Road, Piscataway, NJ 08854, USA}

\author[0000-0002-2838-9033]{Aaron Smith}
\email{asmith@utdallas.edu}
\affiliation{Department of Physics, The University of Texas at Dallas, Richardson, Texas 75080, USA}

\begin{abstract}
Observations with the \textit{JWST} in lensed fields have revealed that galaxies at cosmic dawn may concentrate their star formation in highly dense, compact, star clusters. The high columns and low metallicities encountered in their birth environments suggest that Lyman-alpha (Ly$\alpha$) radiation pressure may be crucial to their formation and evolution. In this study, we address this question by post-processing snapshots from radiation hydrodynamic simulations of dense star cluster-forming clouds ($\Sigma_* \gtrsim 10^3~{\rm M}_\odot~\mathrm{pc}^{-2}$) with a range of dust abundances ($Z_{\rm d}= 0$--$0.1 \, {\rm Z}_{\rm d,\sun}$) using the \textsc{colt} Monte Carlo code. We infer that $\Lya$ is likely to have mild ($\sim 10\%$) effects on the gas-to-star conversion efficiencies ($\epsilon_* \gtrsim 60\%$) for $Z_{\rm d} \gtrsim 0.01 \, {\rm Z}_{\rm d,\sun}$, and even in dust-free environments, $ \epsilon_* \gtrsim25\%$---much higher than the $\lesssim 10\%$ values typical of star-forming regions in the local Universe. This is because the densest filaments dominating stellar mass assembly ($n \gtrsim 10^4~\mathrm{cm}^{-3}$) remain sub-Eddington ($\fEdd<1$). On the other hand, the bulk of the gas volume ($n \lesssim 10^3 \pcm$) has $f_{\rm Edd}>1$, with noticeable fractions having $f_{\rm Edd} \gtrsim 10$, implying that $\Lya$ can launch dynamically significant winds from these systems rapidly ($\lesssim 4 \, \rm Myr$), with possible implications for ionizing photon escape and galactic outflows. The Ly$\alpha$ force multiplier $M_{\rm F}$ is highly sensitive to $Z_{\rm d}$, with $M_{\rm F} \lesssim 3$ ($\lesssim 500$) for $0.1 \, {\rm Z}_{\rm d,\sun}$ (dust-free) environments respectively. Nevertheless, Ly$\alpha$ dominates over UV and IR radiation pressure at all values of $Z_{\rm d} \lesssim 0.1 \, {\rm Z}_{\rm d,\sun}$, by factors of $\sim 3$--$500$. Our results suggest that Ly$\alpha$ radiation pressure reinforces the emerging picture of locally efficient, bursty star formation accompanied by rapid outflows in galaxies at cosmic dawn.
\end{abstract}


\section{Introduction} 

The epoch of cosmic dawn ($z \gtrsim 6$) marks the phase of cosmic history when galaxies have begun assembling stars and black holes within their host dark matter halos, which subsequently drive the reionization of the intergalactic medium. The launch of the \textit{James Webb Space Telescope} \citep[\textit{JWST};][]{Gardner_2023} has revolutionized our view of this epoch, revealing galaxies that are highly compact \citep{Casey_2023,Morishita_2025}, dense \citep{Isobe_2023,Abdurrouf_2024,Topping_2025}, and metal/dust-poor \citep{Nakajima_2023,Curti_2023,Stanton_2026}, with resolved views offered by gravitationally-lensed fields indicating that the bulk of their star formation is concentrated in massive, compact stellar clusters with stellar surface densities ($\Sigma_*$) 2--3 orders of magnitude higher than those in the local Universe \citep{Pascale_2023,Mowla_2024,Adamo_2024,Messa_2026,Claeyssens_2026}. The extreme physical conditions realized in the interstellar medium (ISM) of these galaxies has emerged as a stress-test for our understanding of star and galaxy formation regulated by stellar/AGN feedback across cosmic time. 

The accumulated observational evidence since the launch of the \textit{JWST} points to possible tensions with pre-launch predictions. For instance, the discovery of an overabundant population of UV-bright galaxies at $z\gtrsim 10$, with respect to empirical and model-based predictions, suggests modifications to the mapping between emergent luminosity and dark matter halo masses at these epochs \citep[e.g.,][]{Finkelstein_2023,Castellano_2024,Carniani_2025,Naidu_2026}. In addition, this UV-bright population is characterized by steep UV slopes reflective of very low dust attenuation and gas columns \citep{Saxena_2024,Stark_2026,Tang_2026}, possibly facilitating the leakage of UV photons producing surprisingly early localized ionized bubbles \citep{Witstok_2025,Navarro_2025,McClymont_2025,Marques-Chaves_2026}. Most of the proposed explanations to facilitate these outcomes rely on modifications of the interplay between star formation and feedback in galaxies at this epoch: either (\textit{i}) an elevated baryonic conversion efficiency of gas to stars in halos \citep{Dekel_2023,Somerville_2025}, (\textit{ii}) short-lived bursts of star formation in halos \citep{Shen_2023,Sun_2023,Kravtsov_2024,Gelli_2024,Munoz_2026}, also with potentially elevated star formation efficiencies \citep{Shuntov_2025,Shen_2026,Kar_2026}, (\textit{iii}) limited dust attenuation facilitated by promptly driven outflows by radiation pressure on dust \citep{Ferrara_2023a,Ferrara_2023b} and/or supernovae \citep{Martinez_2026}, and/or (\textit{iv}) a top-heavy initial mass function of stars\footnote{Alternative explanations (directly) unrelated to the baryonic feedback-star formation interplay include (\textit{i}) contributions from accreting black holes, (\textit{ii}) alternative dust grain size distributions or reduced effective core-collapse supernovae yields, or (\textit{iii}) modifications to cosmological structure formation.} \citep{Inayoshi_2022,Yung_2023,Trinca_2024}.

These outcomes at galactic scales are intricately linked to the formation of, and the feedback emerging from, the dense stellar clusters that dominate their rest-frame UV/optical light \citep{Claeyssens_2026,Nakane_2026}. For instance, numerical simulations of star cluster formation have found that the integrated star formation efficiency, i.e. the total fraction of gas converted to stars, is $\gtrsim 50\%$ for the high $\Sigma_*$ conditions realized at cosmic dawn \citep{Grudic_2018,Kim_2018,He_2019,Lancaster_2021c,Fukushima_2021,Menon_2023,Polak_2023,Menon_2025}, significantly higher than the $\lesssim 10\%$ values predicted and observed for star-forming regions in main sequence galaxies in the local Universe \citep{Chevance_23,Pathak_2025}. This finding can be interpreted through the lens of simple Eddington-like arguments, where pre-supernova feedback mechanisms\footnote{The extremely short free-fall times ($\sim 1 \, \rm Myr$) in star-forming clouds at cosmic dawn implies that stars form too quickly for supernovae to regulate star formation locally.} have insufficient momentum to compete with the deep gravitational potential wells at $\Sigma_* \gtrsim 10^3 \, \Msolpc$ \citep{Fall_2010,Thompson_Krumholz_2016,Grudic_2018}. Furthermore, the strong turbulence and short dynamical timescales, necessitated by the high densities realized in these systems, facilitate localized gas/dust ejection and prompt clearing of sightlines for UV photon escape \citep{Crespo_2025,Rivera-Thorsen_2026}, as also suggested by observations of super star cluster-forming regions in local starbursts \citep{Komarova_2024,Beck_2025} and in high-redshift lensed fields \citep{Rivera_2019,Pascale_2023,Kim_2023_escape}. There are also possible observational signposts of an excess of massive stars (i.e. a top-heavy IMF) in conditions of high $\Sigma_*$ \citep{Schneider_2018}; \citet{Menon_2024} shows that even if this were the case, star formation would remain relatively efficient, with feedback only acting to eject gas much more promptly and vigorously. Overall, these studies suggest that the clues to the distinct outcomes inferred for galaxies at cosmic dawn may lie in the altered feedback--gravity competition of their high $\Sigma_*$ star-forming clumps. Indeed, semi-analytic galaxy formation models incorporating these bottom-up insights have yielded UV luminosity and stellar mass functions that are in better agreement with observations than pre-\textit{JWST} models at $z\gtrsim 10$ \citep{Somerville_2025}. 

However, none of these studies fully consider the contribution of $\Lya$ radiation pressure. The oversight is related to its high dimensionality and complexity, as well as the inherently demanding computational requirements for modeling it self-consistently. Since $\Lya$ photons have extremely high cross-sections to abundantly present hydrogen atoms, they undergo resonant scattering, and impart a cumulative radially-outward momentum that can drastically exceed the energy carried by each photon \citep{Adams_1975,Neufeld_1990,Bithell_1990,dijkstra_ly-driven_2008}. These considerations have motivated semi-analytic and Monte Carlo radiative transfer calculations quantifying the momentum injection by $\Lya$ radiation pressure in idealized slab/spherical cloud geometries and source distributions \citep{Smith_2017, michel-dansac_rascas_2020, Lao_2022, smith_lyman_2025}, finding that it can dominate over the momentum input from other pre-supernovae mechanisms by factors of $\sim 10$--$1000$ \citep{kimm_impact_2018,nebrin_lyman-_2025}, depending sensitively on the dust optical depth at $\Lya$ wavelengths \citep{tomaselli_lyman-alpha_2021}. Attempts at incorporating these insights into subgrid models for $\Lya$ feedback in (low-mass/metallicity) simulations \citep{kimm_impact_2018,Han_2022} and semi-analytic models \citep{Nebrin_2023,Kapoor_2023} reveal that the formation of Population III stars and star clusters may be significantly impacted by including Ly$\alpha$ feedback. Most recently, \citet{nebrin_lyman-_2025} developed novel semi-analytic calculations to quantify the role of $\Lya$ radiation pressure, incorporating a wide range of effects---notably including continuum dust absorption, gas velocity gradients, and turbulent density distributions---all of which could conspire to weaken the effects of $\Lya$. Despite these considerations, the authors find that $\Lya$ radiation pressure is still capable of dominating other early feedback mechanisms, even at dust abundances approaching the solar neighborhood value. Studies building  on these analytical results with 1D semi-analytic shell models \citep{Ferrara_2025,manzoni_lyman-_2025} estimate that $\Lya$ radiation pressure precludes efficient star formation ($\gtrsim 50\%$) in dense star clusters except for the most extreme cases ($\Sigma_* \gtrsim 10^5 \, \Msolpc$) and/or near-solar dust abundances. Collectively, these results introduce complications for interpretations of \textit{JWST} observations that invoke elevated star formation efficiencies.

It is important to note that neutral hydrogen (\HI) gas in star-forming clouds are highly turbulent and filamentary, with the sources of stellar radiation and the ionized hydrogen (\HII) regions they form being distributed in space rather than point sources---complications that cannot be accounted for fully in semi-analytic calculations, yet are likely to have implications for $\Lya$ momentum injection. For instance, in supersonic turbulent media across scales, stars form and accrete through filaments with densities/columns that exceed the average in the cloud by factors $\sim 10$--$100$, remaining relatively unaffected by feedback as gravity dominates in them; however, lower density columns are susceptible to ejection by radiation pressure, with the transition column where gravity wins out being dependent on the momentum injection rate per unit stellar mass \citep{Thompson_Krumholz_2016}. This explains why 3D numerical simulations quantifying the impact of radiation pressure from stellar UV and dust-reprocessed IR photons on dusty gas reveal that the outcomes can depart from simple analytical expectations \citep{Krumholz_Thompson_2012,Krumholz_Thompson_2013,Rosen_2016,Raskutti_2017,Kim_2018,Menon_2023}. These differences are aggravated by additional 3D effects that diminish the impact of radiation pressure: such as the tendency for the photon flux to preferentially be higher in low density gas (i.e. the matter--radiation anticorrelation) \citep{Skinner_2015,Menon_2022b}, and the vector cancellation of fluxes in stellar populations with multiple radiation sources \citep{Kim_2018,Menon_2023}. It remains to be seen to what extent the aforementioned impacts of $\Lya$ radiation pressure in regulating star formation and driving winds are affected by similar mechanisms in realistic star-forming environments. 

In this paper, we address this question directly by computing the $\Lya$ radiation pressure force in post-processing on snapshots from a suite of radiation-hydrodynamic (RHD) simulations of the formation of dense star clusters from turbulent, self-gravitating clouds \citep{Menon_2025}, using the Monte Carlo $\Lya$ radiative transfer code \textsc{colt} \citep{smith_lyman_2015,mcclymont_modelling_2025}. We span a range of dust abundances representative of metal-poor environments at the epoch of reionization ($Z_{\rm d} \sim 0$--$0.1 \, {\rm Z}_{\rm d,\sun}$) allowing us to quantify the sensitivity of $\Lya$ feedback on dust opacity in a fully 3D, turbulent medium with radiation sources formed self-consistently through turbulent fragmentation. We focus on two central questions: (\textit{i})~whether $\Lya$ radiation pressure can prevent efficient star formation during the assembly of dense star clusters at cosmic dawn, and (\textit{ii})~whether it can drive dynamically significant winds once star formation has saturated. The remainder of this paper is organized as follows. In Section~\ref{sec:methods} we describe our RHD simulations, the $\Lya$ radiative transfer post-processing, and our treatment of dust. Section~\ref{sec:results} presents our results for the $\Lya$ force-gravity competition in our simulations, and the momentum budget relative to other radiative feedback mechanisms. We discuss broader implications in Section~\ref{sec:discussion} and summarize our conclusions in Section~\ref{sec:conclusions}. Finally in Appendix~\ref{sec:Appendix_fEdd} and ~\ref{sec:Appendix_numerical} we quantify the sensitivity of our results to our analysis approach and numerical choices. 

\section{Methods}
\label{sec:methods}

\subsection{Simulations of Dense Star Cluster Formation}

For this work, we use the gas and stellar properties obtained with radiation hydrodynamic simulations of the formation of a young star clusters from turbulent, self-gravitating clouds. The simulation setups are  similar to that described in \citet{Menon_2025}, with additional physics implemented and different choices for the subgrid stellar population as described below. We briefly summarize the most relevant aspects, focusing on the differences, and refer the reader to the earlier works for more detailed information. 

The simulations are initialized with a turbulent, self-gravitating molecular cloud of mass $10^6 \, \Msun$ and radius $10 \, \rm pc$, corresponding to a surface density of $\sim 3000 \, \text{M}_{\sun} \, \rm pc^{-2}$. The cloud is evolved in a adaptive mesh refinement (AMR) domain of size $40 \, \rm pc \times 40 \, \rm pc \times 40 \, \rm pc$, with a maximum effective resolution of $1024^3$ cells (minimum cell size $\sim 0.03 \, \rm pc$), with regions outside the cloud a 100 times less dense. Turbulence is driven using the method described in \citet{Federrath_2010,Federrath_2022}, to achieve a velocity dispersion $\sigma_{\rm v} \sim 23 \, \rm km/s$ sufficient for the cloud to be marginally stable ($\alpha_{\rm vir} = 2 E_{\rm kin}/E_{\rm grav} = 2$), with a power spectrum $E(k) \propto k^{-2}$ containing a natural mixture of solenoidal and compressive modes. The turbulence is driven for 5 eddy turnover times ($5 L/\sigma_{\rm v}$) without self-gravity to achieve a steady state for the turbulence+thermochemistry, using the \texttt{TURBSPHERE} approach of using a static potential to confine the gas in this phase \citep{Lane_2022}. This serves as the initial conditions for the subsequent phase of turbulent collapse where self-gravity is turned on. Unresolved star formation is represented by sink particles using the implementation described in \citet{Federrath_2010_Sinks,Federrath_2011}, which we interpret as unresolved subclusters; for the purpose of assigning radiative feedback, each sink is assumed to be fully sampled in the IMF. We compute the corresponding stellar radiative output with the BPASS v2.2 model \citep{Eldridge_2017,Stanway_2018}.

Radiation hydrodynamics is performed with the Variable Eddington Tensor--closed \texttt{VETTAM} implementation \citep{Menon_2022} in the \texttt{FLASH} code \citep{Fryxell_2000,Dubey_2008}. We solve the radiative transfer in four bands (Lyman continuum, Lyman--Werner, photoelectric, and a gray infrared band), which together provide the heating rates, and photoionization and photodissociation rates for different chemical species and dust, with dust cooling via IR emission treated self-consistently in the IR band assuming dynamical coupling to the gas \citep{Menon_2022b}. These rates are included in a minimal chemical network \citep[i.e. as described in][]{Kim_2023} using the implementation described in an upcoming paper (Menon et al., 2026, in prep). The chemical network solves for the non-equilibrium abundances of H, $\rm H_2$, $\rm H^+$ and $\rm e^{-}$, and adopts a steady-state approximation for the C/C+ ionization balance, and charge transfer dictated O/O+ abundances; the CO abundance is based on fits to PDR models of \citet{Gong_2017}. While the network is highly idealized, we find that it can capture the thermal evolution of the gas as well as an extensive chemical network, and therefore suffices for the purposes of this study. We refer the reader to earlier work \citep{Menon_2023,Menon_2025} for the gas/dust opacities and numerical choices for evolving the hydrodynamics/radiation transport. We do not consider any non-radiative feedback mechanisms in the simulations (e.g. stellar winds or protostellar jets) nor do we include magnetic fields, and the dust dynamics are coupled to the gas with opacities assuming a spatially- and temporally-fixed grain size distribution \citep{Weingartner_Draine_2001a}.

\subsection{Post-processing $\Lya$ Radiative Transfer}

We post-process snapshots of our RHD simulations with the Cosmic $\Lya$ Transfer (\textsc{colt}) code \citep{smith_lyman_2015, smith_physics_2019, smith_physics_2022, mcclymont_modelling_2025}. We construct the gas distribution in \textsc{colt} using the simulated gas densities and temperatures, and we use the positions, masses, and ages of sink particles to define the sources of UV radiation. We neglect any contribution from a intragalactic and metagalactic UV background, which is expected to be subdominant compared to the radiation field produced by the massive clusters in our simulated volumes. For each sink, we compute the emergent source spectrum with BPASS, consistent with the stellar-population model used in the on-the-fly RHD calculations. We however, increase the spectral resolution well beyond the three discrete bands captured on-the-fly in the simulations.

We then use \textsc{colt} to compute the ionization state of H, He, and metals by propagating $10^8$ Monte Carlo (MC) photon packets, iterating until the total $\mathrm{H}^+$ recombination rate in the simulation volume is converged to $0.1\%$. Although our RHD simulations evolve H ionization on the fly, we adopt the \textsc{colt} solution for the subsequent $\Lya$ transfer to be more consistent, and as the Monte Carlo radiative transfer (MCRT) approach in post-processing can potentially achieve higher angular and spectral resolution. That being said, we find that the ionization states with VETTAM are in good agreement with \textsc{colt} for the spatial structure of \HII regions and the distribution of H ionization fractions. 

Given the converged ionization structure, \textsc{colt} performs a ray-based transfer of individual $\Lya$ photon packets through the medium, including contributions from (\textit{i}) radiative recombination emission, and (\textit{ii}) collisionally excited $\Lya$ emission. The code follows resonant scattering in neutral hydrogen and includes absorption and scattering by dust. For the purposes of this paper, our primary diagnostic is the momentum imparted to the gas by $\Lya$ photons. We estimate the local momentum imparted to each gas cell by computing a vector-sum over the momentum increments deposited by the ensemble of photon packets undergoing scattering events within that cell. We note that this converges faster than path-based momentum estimators. We have also checked that a diffusion-based force estimator calculated from the gradient of the $\Lya$ energy density agrees with the scattering-based approach in regimes where we expect the diffusion approximation to apply (i.e. at early times when optically-thin channels are rare).

To reduce the computational cost of the resonant-scattering random walk, we employ moderate core skipping, i.e. a random excursion into the wings of the line facilitated by an interaction with an atom of sufficient velocity, preventing repetitive scatterings with negligible spatial diffusion \citep{ahn_ly_2002}. While this technique can significantly speed up calculations without affecting emergent $\Lya$ photon statistics/profiles, it might have implications on the $\Lya$ radiation pressure force which relies on trapping in the core and wing for effective momentum deposition and high force multipliers \citep{Lorinc_2025}. An important numerical choice relevant to core skipping is $\xcrit$, which is in units of the dimensionless Doppler frequency $x \equiv (\nu - \nu_0) / \Delta \nu_{\rm D}$ for a given thermal Doppler width $\Delta \nu_{\rm D} \equiv (v_{\rm th}/c)\,\nu_0$ and thermal velocity $v_{\rm th} \equiv (2 k_\text{B} T / m_\text{H})^{1/2}$, beyond which photons are pushed out to facilitate escape from the core. Since the photon optical depth and therefore the number of scatterings drop further away from the core, higher values of $\xcrit$ will be computationally faster but represent the internal radiation field less accurately. We adopt a constant value of $\xcrit = 2$ in our calculations as a conservative balance between computational efficiency and physical accuracy. Another numerical choice of relevance here is the number of photon packets launched ($n_{\rm ph}$), which is important for ensuring convergence of the $\Lya$ radiation field, especially at large optical depths \citep{Camps_2018}. We adopt a total of $n_{\rm ph}=10^6$ photon packets in this study. In Appendix~\ref{sec:Appendix_numerical}, we quantify the dependence of our results on these choices by repeating the calculations for $\xcrit \in \{1,2,3\}$ and $n_{\rm ph} \in \{10^5,10^6,10^7\}$, and demonstrate that the conclusions of this study are insensitive to these choices.

\subsection{Treatment of Dust}

Dust can act as a source of continuum opacity for $\Lya$ photons, destroying $\Lya$ photons and thereby setting an upper limit on the $\Lya$ radiation pressure force \citep[e.g.,][]{Oh_2002,tomaselli_lyman-alpha_2021,nebrin_lyman-_2025}. The choice of dust opacity used in the \textsc{colt} $\Lya$ radiation transport step is therefore likely to be critical to the inferred dynamical impact of $\Lya$ radiation pressure. In the simulations, we assume a uniform dust-to-gas ratio ($Z_{\rm d}$) that scales linearly with the initial metallicity ($Z$) and does not evolve; i.e. we do not include metal enrichment or dust injection, so both $Z$ and $Z_{\rm d}$ are fixed in space and time. We focus on two metallicities---$0.1\,\text{Z}_{\sun}$ and $0.01\,\text{Z}_{\sun}$---chosen to probe the parameter regime relevant for observed galaxies with the \textit{JWST} at the Epoch of Reionization. We adopt values of $Z_{\sun} = 0.0134$ and ${\rm Z}_{\rm d, \sun} = 1/162$ in this work and the \citet{Weingartner_Draine_2001a} grain optical properties using the Small Magellanic Cloud (SMC) dust size distribution model.  We repeat these choices for their corresponding \textsc{colt} post-processing calculations. 

In addition to these, we repeat the \textsc{colt} post-processing steps (ionization and $\Lya$ transport) for the $0.01\,\text{Z}_{\sun}$ simulation outputs, but with (\textit{i}) an assumed dust abundance of $4 \times 10^{-4}\,\text{Z}_{\sun}$ and (\textit{ii}) a dust-free case\footnote{We note that this represents an inconsistency in the assumed $Z_{\rm d}$ in the RHD simulation and the \textsc{colt} post-processing step. However, we do not expect any noticeable impacts of this inconsistency on our results, since the star formation histories and gas density distributions are relatively insensitive to $Z$ \citep{Menon_2025}.}. The former case is motivated by empirical findings of a super-linear relation between $Z$ and $Z_{\rm d}$ at $Z \lesssim 0.1\,\text{Z}_{\sun}$ \citep{remy-ruyer:2014}, as the the empirical fit for $Z = 0.01\,\text{Z}_{\sun}$ corresponds to $Z_{\rm d} = 4 \times 10^{-4} \, {\rm Z}_{\rm d,\sun}$. However, as we will discuss in Section~\ref{sec:dust_discussion}, it is the dust opacity at $\Lya$ wavelengths that is relevant here, and an alternate grain size distribution can correspond to this range of values even at higher $Z$. On the other hand, the dust-free case represents an upper limit on the impact of $\Lya$ radiation pressure.

\section{Results}
\label{sec:results}

\begin{figure*}
  \includegraphics[width=\linewidth]{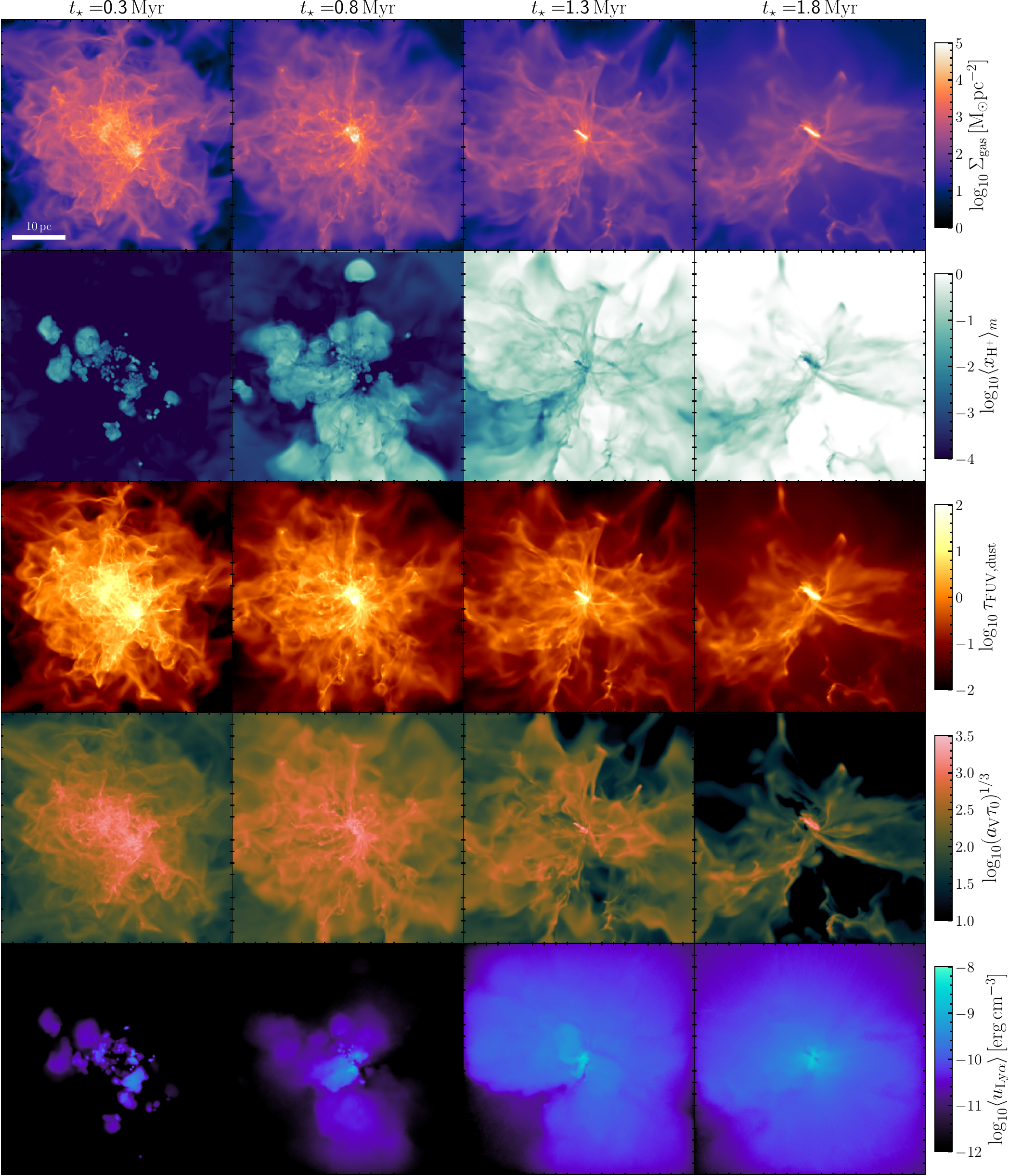}
  \caption{Time evolution of relevant quantities in the simulation, shown at representative times spanning the early embedded phase through the later, feedback-dominated evolution. Each column corresponds to a different snapshot time (increasing from left to right). From top to bottom we show: the projected gas surface density, $\Sigma_{\rm gas}$; the mass-weighted projected average of the hydrogen ionization fraction, $\langle x_{\rm H^+} \rangle_m$; the dust optical depth to far-UV photons, $\tau_{\rm FUV,dust}$, computed using the dust opacity adopted in the RHD simulations ($\kappa_{\rm d,PE}\sim 500\,{\rm cm^2\,g^{-1}}$); a normalized version of the line-center optical depth to $\Lya$ photons, $(a_{\rm V}\tau_{0})^{1/3}$ where $a_{\rm V}$ is the Voigt parameter; and $\langle u_{\Lya} \rangle$ the volume-weighted average $\Lya$ radiation energy density.}
  \label{fig:Snapshots}
\end{figure*}

\begin{figure*}
  \includegraphics[width=\linewidth]{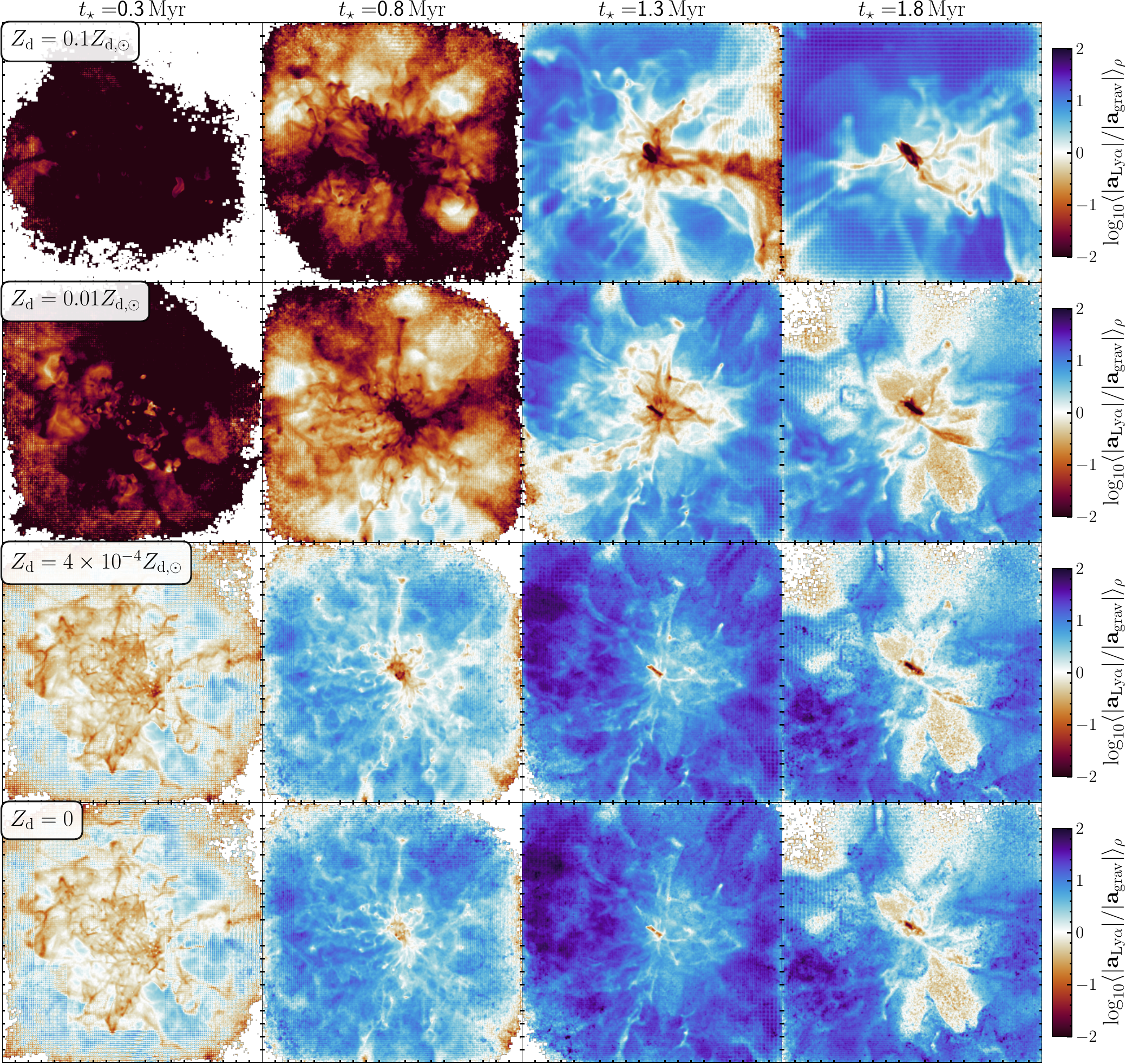}
  \caption{Mass-weighted projections of the ratio of the magnitude of the acceleration due to $\Lya$ radiation pressure ($a_{\Lya}$) and due to the stellar and self-gravitational potential ($a_{\rm grav}$) at different times (columns), and for the four different dust abundances explored in this study (rows). We can clearly see that over time, as gas gets consumed by star formation and ejected by radiative feedback physics included in the simulation (i.e. other than $\Lya$), more of the gas becomes super-Eddington, with the extent depending strongly on dust abundance. We can also see a clear correlation with gas density, with the dense compact central region and the filamentary flows feeding it remaining sub-Eddington even at later times due to their higher $\agrav$ and $\tau_{\rm FUV,dust}$.}
  \label{fig:fEddSnapshots}
\end{figure*}

\subsection{Overall evolution}

We begin by describing the overall time evolution of the simulations, with particular emphasis on the growth of stellar mass, the resulting H\,\textsc{ii} regions, and the morphology of the gas through which $\Lya$ photons undergo resonant scattering. The evolution is broadly similar to that in \citet{Menon_2025}; we refer the reader to that work for additional details. In Figure~\ref{fig:Snapshots} we show the evolution of several relevant quantities for the $0.01\,\text{Z}_{\sun}$ simulation, at times representative of different evolutionary stages. We show, in different rows, (\textit{i}) the total projected gas surface density $\Sigma_{\rm gas}$, (\textit{ii}) the mass-weighted average ionization fraction of hydrogen $\langle x_{\rm H^+} \rangle_m$, (\textit{iii}) the optical depth of dust to FUV photons $\tau_{\rm FUV,dust}$,\footnote{We use the dust opacity for the photoelectric heating UV band (6--11.2\,eV) used in our simulations ($\kappa_{\rm d,PE} \sim 500 \, \rm cm^2 g^{-1}$) that overlaps with the $\Lya$ line for this plot.} (\textit{iv}) $(a_{\rm V} \tau_0)^{1/3}$, a normalized representation of the line-center ($\nu_0$) $\Lya$ optical depth ($\tau_0$), and (\textit{v}) $\langle u_{\Lya} \rangle$ the volume-weighted average $\Lya$ radiation energy density. For $\tau_0$, we use the expression for the $\Lya$ cross-section
\begin{equation}
    \sigma(\nu) = \sigma_{0} \phi_{\rm Voigt} (\nu) \, ,
\end{equation}
where
\begin{equation}
    \label{eq:Lya_crosssection}
     \sigma_{0} = 5.898 \times 10^{-14} \, (T/10^4 \, \rm K)^{-1/2} \, \rm cm^2 \, ,
\end{equation}
is the cross-section at line center and $\phi_{\rm Voigt} (\nu)$ is the Voigt profile \citep{Voigt1912}. We use the Voigt parameter for $\Lya$, $a_{\rm V} = 4.7 \times 10^{-4} \, (T/10^4 \rm K)^{-1/2}$ for the quantity $(a_{\rm V}\tau_{0})^{1/3}$, which analytical models predict corresponds to the $\Lya$ force multiplier ($M_{\rm F}$) for a dust-free uniform medium \citep{Adams_1975,Bonilha_1979,lao_resonant-line_2020}. However this is an order of magnitude estimate for force multiplication and is simply shown for a visual reference of the relative changes with time in Figure \ref{fig:Snapshots}. 

At early times, turbulent fluctuations seeded in the cloud grow through self-gravity and form filamentary structures. Portions of these filaments collapse to form the first sink particles (representing subclusters of stars), which continue to grow via accretion along the filaments until their UV luminosities are sufficient to produce the first compact H\,\textsc{ii} regions (Figure~\ref{fig:Snapshots}, first column). In spite of forming \HII regions the stellar mass continues to grow due to anisotropic accretion and the formation of new fragments. The associated increase in the ionizing flux facilitates the growth of the H\,\textsc{ii} regions, eventually merging with neighboring ones and, in some cases, forming density-bounded channels (Figure~\ref{fig:Snapshots}, second column). The distribution of \HII regions and $\Lya$ radiation are still relatively compact at these times. Over time, the regions ultimately merge into a cluster-scale ionized nebula that, however, still exhibit substantial gas/dust columns\footnote{We do not destroy dust or change the dust-to-gas ratio in the ionized gas unless dust temperatures approach the sublimation temperature ($\sim 1200 \, \rm K$) and/or dust is destroyed by thermal sputtering, which occurs at gas temperatures above those typical of photoionized gas ($>10^6 \, \rm K$).} because the ionized gas densities remain high within the \HII regions (Figure~\ref{fig:Snapshots}, third row). Finally, radiation pressure on dust facilitates further expansion of the gas, reducing the dust columns and increasingly evacuating most the region, except a dense flattened structure of gas that occupies a small fraction of the solid angle around the cluster (Figure~\ref{fig:Snapshots}, fourth row). 

\subsection{Lyman-$\alpha$ radiation pressure-gravity competition}

In Figure~\ref{fig:Snapshots} we showed that the value of $(a_{\rm V} \tau_0)^{1/3}$ can be extremely high ($\gtrsim 10^3$), especially at early times when the star cluster is being assembled. Since the amount of $\Lya$ radiation pressure scales with this quantity, these values suggests that the momentum from resonant line scattering can contribute significantly to the competition of feedback and gravity. To quantify this, we use the (net) acceleration from radiation pressure by $\Lya$ photons calculated by \textsc{colt} in each computational cell ($\aLya$), computed from the vector sum of the individual MC photon scattering events occuring in it. This net acceleration is compared to the combined gravitational acceleration from the stellar sources and the self-gravity ($\agrav$). The former is simply a direct sum over the point sources with a softening of $\sim 0.07 \, \rm pc$, and the latter is computed from the gradient of the (self-) gravitational potential through a centered-differencing scheme. We compare the accelerations by calculating the Eddington ratio ($\fEdd$) given by
\begin{equation}
\label{eq:fEdd}
    \fEdd = \frac{|\aLya \cdot -{\agrav}|}{|\agrav|^2} \, ,
\end{equation}
where we compare the component of the radiative acceleration in direct (local) opposition to gravity. We also experimented with other approaches to computing the \textit{local} Eddington ratio: in Appendix~\ref{sec:Appendix_fEdd} we compare the distributions of $\fEdd$ obtained by Equation~(\ref{eq:fEdd}) with these alternatives and demonstrate that they would not change the conclusions of this study. We also compute a \textit{global} Eddington ratio in Appendix~\ref{sec:Appendix_fEdd}, i.e., considering the star cluster as an effective single source placed at its center of mass and quantifying the integrated ratio of radiative and gravitational forces seen by this source. While the integrated nature of this estimator is simple to interpret and robust to Monte Carlo noise, it ignores the nuances implied by the broad density distribution in the gas and the local competition between feedback and gravity that depends on the density, which we will show is a critical aspect to consider for this problem.

In Figure~\ref{fig:fEddSnapshots} we show the distribution of the mass-weighted projection of $\fEdd$ for all four dust cases considered in this work of Figure~\ref{fig:Snapshots}, and the representative times sampled in Figure~\ref{fig:Snapshots}. There is clearly a sensitive dependence of $\fEdd$ on the assumed value of $Z_{\rm d}$, indicating that dust absorption clearly limits the dynamical impact of $\Lya$ radiation pressure. We can also see that there is a common time evolution of $\fEdd$, with increasing fraction of super-Eddington regions at later times, as the stellar mass increases and the amount of gas decreases due to consumption and feedback mechanisms (other than $\Lya$). Furthermore, there is also a tendency for central regions that contain the densest gas to have lower $\fEdd$ and vice-versa. Below, we quantify the implications of these qualitative findings in further detail. 

\subsubsection{Effect on the Efficiency of Star Formation}

\begin{figure*}
  \centering
  \includegraphics[width=\textwidth]{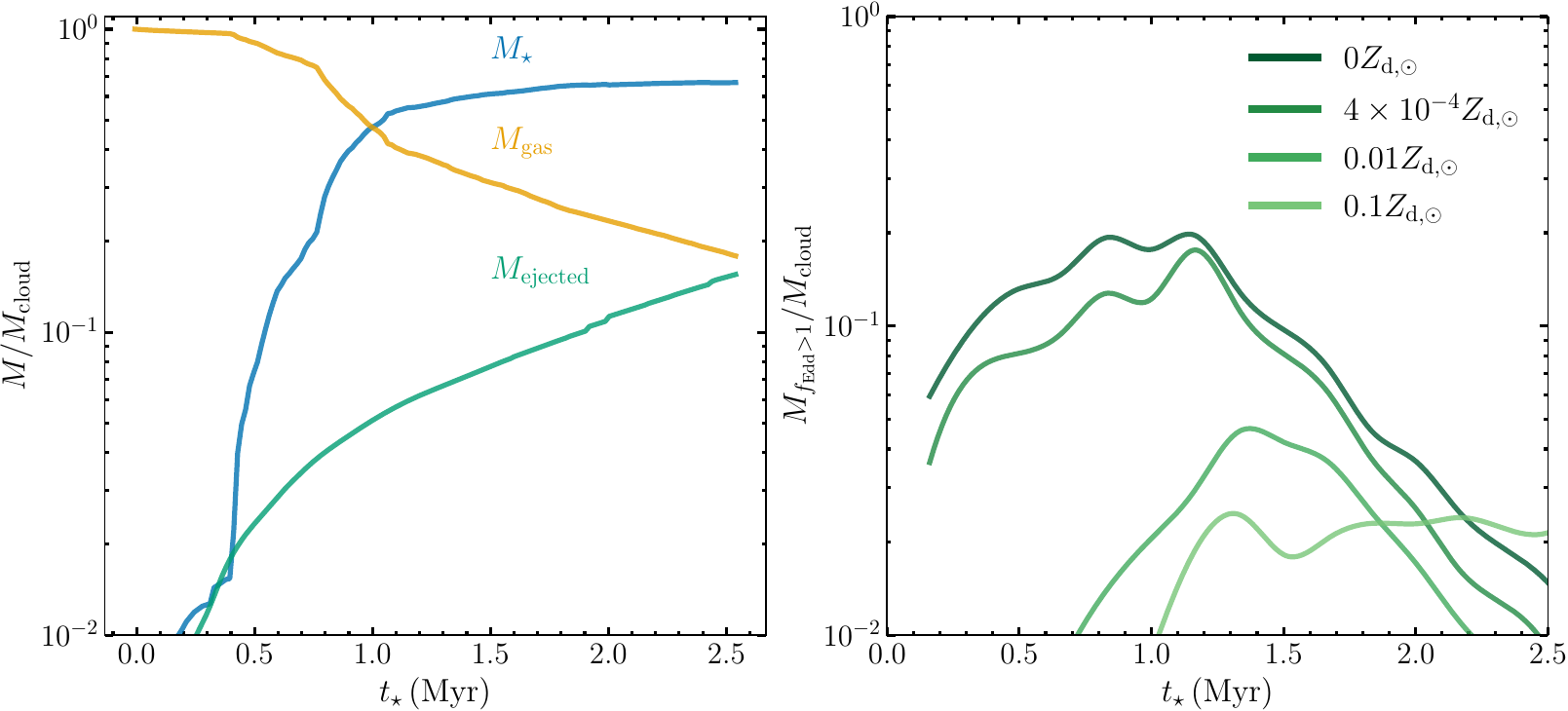}
  \caption{The potential impact of $\Lya$ radiation pressure on the global competition between feedback and star formation. \textbf{Left}: Lines show the total stellar mass formed ($M_\star$; blue), remaining gas mass ($M_{\rm gas}$; orange), and the mass ejected by radiative feedback physics included in the simulation ($M_{\rm ejected}$; green), all normalized to the initial cloud mass ($\Mcloud = 10^6 \, \Msun$), as a function of the age of the stellar population ($t_\star$). \textbf{Right}: The fraction of $\Mcloud$ which is super-Eddington (i.e. $\fEdd>1$) to $\Lya$ radiation pressure as a function of time, for the four different dust cases considered in this work. We can see that $M_{\fEdd>1}/\Mcloud$ is low for all cases at early times when stellar mass is being assembled, and only peaks once $M_\star$ has achieved values $\gtrsim 50\%$, suggesting that $\Lya$ radiation pressure is unlikely to preclude efficient star formation in these systems. Beyond this point, however, $\Lya$ is likely to eject a significant fraction of the mass---especially at low dust abundances---and will therefore strongly affect the time evolution of $M_{\rm gas}$ and $M_{\rm ejected}$.}
  \label{fig:fEddMainPlot}
\end{figure*}

To see whether the $\Lya$ radiation pressure could have potentially limited the amount of star formation were it included, we compute the cumulative mass of gas that has $\fEdd>1$ (denoted as $M_{\fEdd>1}$), and compare them to time evolution of the cumulative stellar mass assembled ($M_\star$), the total mass ejected ($M_{\rm ejected}$) from the computational domain (primarily) by radiation pressure on dust and photoionization \citep{Menon_2023}, and the remaining gas mass ($M_{\rm gas}$) mass in the system. The total mass is conserved by the closure $M_\star + M_{\rm gas} + M_{\rm ejected} = \Mcloud$ where $\Mcloud = 10^6\,\Msun$ is the initial mass of the cluster-forming turbulent cloud we model. In Figure~\ref{fig:fEddMainPlot}, we show the time evolution of these quantities (left) and the super-Eddington mass $M_{\fEdd>1}$ (middle) normalized with respect to $\Mcloud$, for all four dust cases. We can clearly see that the cases with $Z_{\rm d} \gtrsim 0.01\,{\rm Z}_{\rm d,\sun}$ are unlikely to have noticeable effects on the integrated star formation efficiency (SFE = $M_\star/\Mcloud$) as they attain a maximum of $M_{\fEdd>1}/\Mcloud \lesssim 5\%$, whereas $M_\star \sim 65\%$ at the end of the simulation. On the other hand, for lower $Z_{\rm d}$, the impact is much more noticeable, with maximum values of $M_{\fEdd>1}/\Mcloud$ of $\sim$ 15\% and 20\% respectively. However, even in these cases, it is critical to note that $M_{\fEdd>1}/\Mcloud$ achieve these maximum values only once the SFE has already saturated. At early times ($t_\star \lesssim 0.7 \, \rm Myr$) when the cloud is collapsing and the stellar mass is still being assembled (low SFE), $M_{\rm gas}/\Mcloud \gtrsim 80\%$, yet $M_{\fEdd>1}/\Mcloud \lesssim 15\%$ even for the dust-free limit, suggesting that $\Lya$ would have been unable to prevent the bulk of the gas mass from forming and/or being accreted onto the star cluster at these early phases of stellar mass assembly. 

\begin{figure}[]
  \centering
  \includegraphics[width=0.5\textwidth]{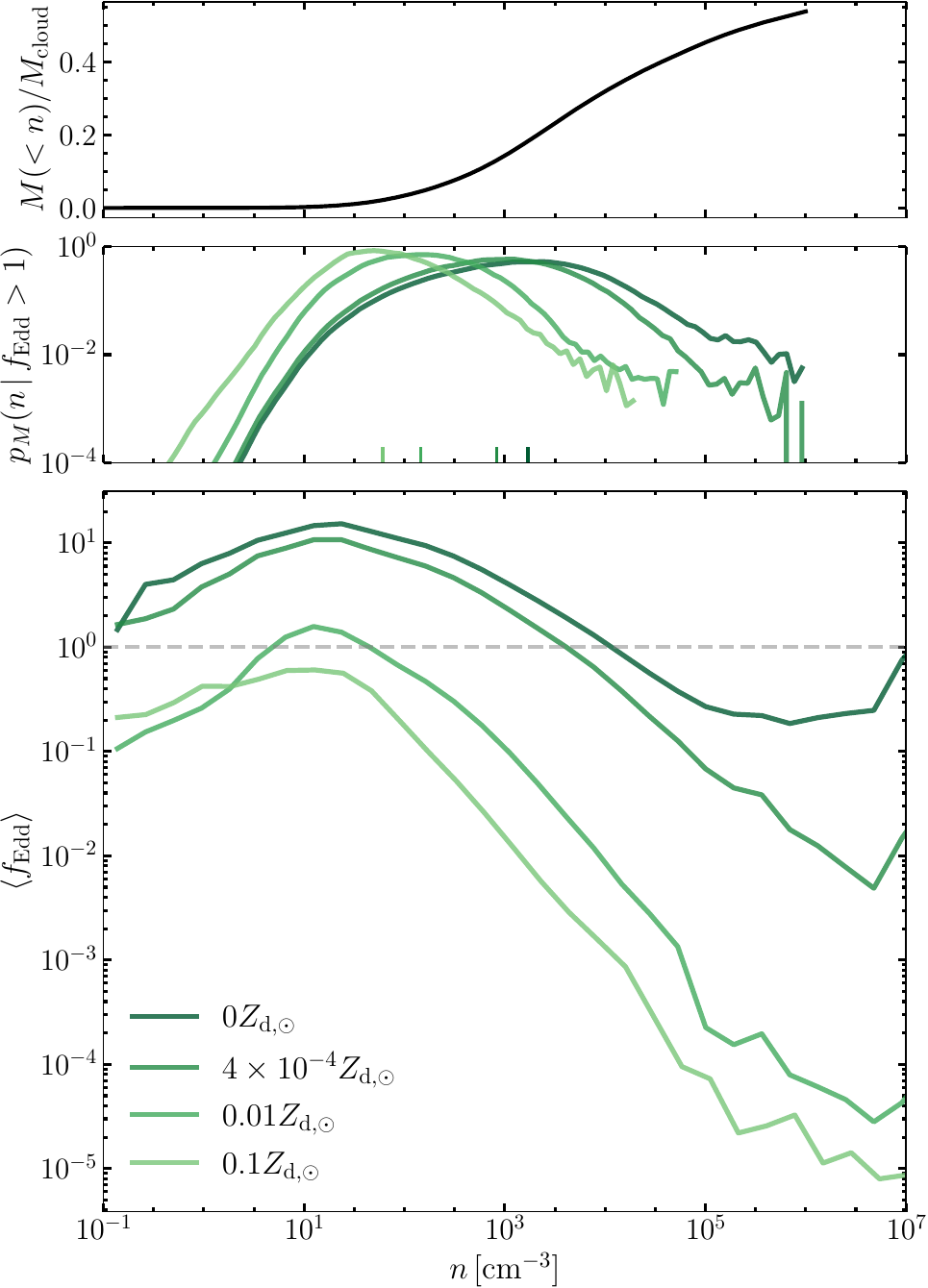}
  \caption{\textbf{Main Panel}: Mean value of the Eddington ratio (Equation~\ref{eq:fEdd}) for gas at a number density ($n$) for the different dust treatments at the time $t_\star = 0.83 \, \rm Myr$. We can see that $\langle \fEdd \rangle$ decreases at the low $n \lesssim 10 \, \pcm$ due to their lower optical depths, and more noticeably so for high $n$ due to its stronger gravity, with higher dust abundance reflecting in a transition from super- to sub-Eddington at lower $n$. \textbf{Top-middle panel}: shows the distribution across $n$ of super-Eddington ($\fEdd>1$) gas, by plotting the mass-weighted PDF of $n|\fEdd>1$, with the colored ticks on its $x$-axis indicating the median. \textbf{Top panel}: indicates the fraction of the initial cloud mass ($M_{\rm cloud}$) in gas with number densities $<n$, which along with the typical $n$ at which $\langle \fEdd \rangle<1$, controls the amount of super-Eddington mass reflected in Figure~\ref{fig:fEddMainPlot}.}
  \label{fig:fEdd_vs_n}
  
\end{figure}

The impact of $\Lya$ at these early times are likely weak as the ionizing flux and $\Lya$ luminosity, which are proportional to $M_\star$, are low, whereas the gravitational potential well (due to $M_{\rm gas}+M_\star$) remains high, explaining why $M_{\fEdd>1}/\Mcloud$ increases with time. Another reason for this is that the gas density in the cloud---especially the densest gas---decreases as it is consumed by star formation and/or ejected by radiative feedback in the simulation. This is pertinent because Figure~\ref{fig:fEddSnapshots} suggests that the densest central regions have lower values of $\fEdd$ than regions further out at lower density. We quantify this relation by plotting $\langle \fEdd \rangle$ the binned average value of $\fEdd$ at a given gas number density $n$ in Figure~\ref{fig:fEdd_vs_n} at a time $t_\star = 0.83 \, \rm Myr$ corresponding to when the dust-free case peaks. At low densities, $\langle \fEdd \rangle$ increases with $n$, consistent with the expectation of more resonant scattering in denser gas. However, after an inflection point there is a clear anti-correlation with $n$ with the gas at the highest $n$ always sub-Eddington for all dust cases. This is also reflected in the (mass-weighted) distribution of $n$ for $\fEdd>1$ gas, showing that the peak and tails of the distribution shifts to higher $n$ for lower $Z_{\rm d}$. The maximum $n$ up to which $\langle \fEdd \rangle >1$ is higher for lower dust abundances: on-average $\lesssim 100 \pcm$ for $0.1\,\text{Z}_{\sun}$ and $\lesssim 10^4 \pcm$ for the dust-free limit, respectively. We note that we have selected this curve at a time ($t_\star = 0.83 \, \rm Myr$) where the SFE ($\sim 45\%$) has almost reached its saturation value, yet, the top panel in Figure~\ref{fig:fEdd_vs_n} indicates that the bulk of the gas mass is at $n$ beyond the super- to sub-Eddington transition point, even for the dust-free limit. This explains why $M_{\fEdd>1}/\Mcloud$ remains relatively low at all times and we expect implications on the SFE to be mild as most of the gas mass and stellar assembly occurs in gas at the highest $n$, where $\fEdd <1$. Inspection of the corresponding curves at other times reveals that the $n$ where this transition occurs is relatively independent of time. Therefore, the increase of $M_{\fEdd>1}/\Mcloud$ in Figure~\ref{fig:fEddMainPlot} is a result of the densest gas being consumed by star formation (and dispersed by radiative feedback), a larger fraction of the instantaneous gas mass becomes super-Eddington. $M_{\fEdd>1}/\Mcloud$ falls at late times as $M_{\rm gas}$ is decreasing, and low-density channels for $\Lya$ escape open up.

Overall, our results suggest that even at very low dust abundances, $\Lya$ radiation pressure is unlikely to preclude efficient star formation ($\gtrsim 50\%$) in dense star cluster-forming sites as the densest gas that dominates stellar mass assembly remains sub-Eddington.

\subsubsection{Effect on mass ejection}

\begin{figure}[]
  \centering
  \includegraphics[width=0.5\textwidth]{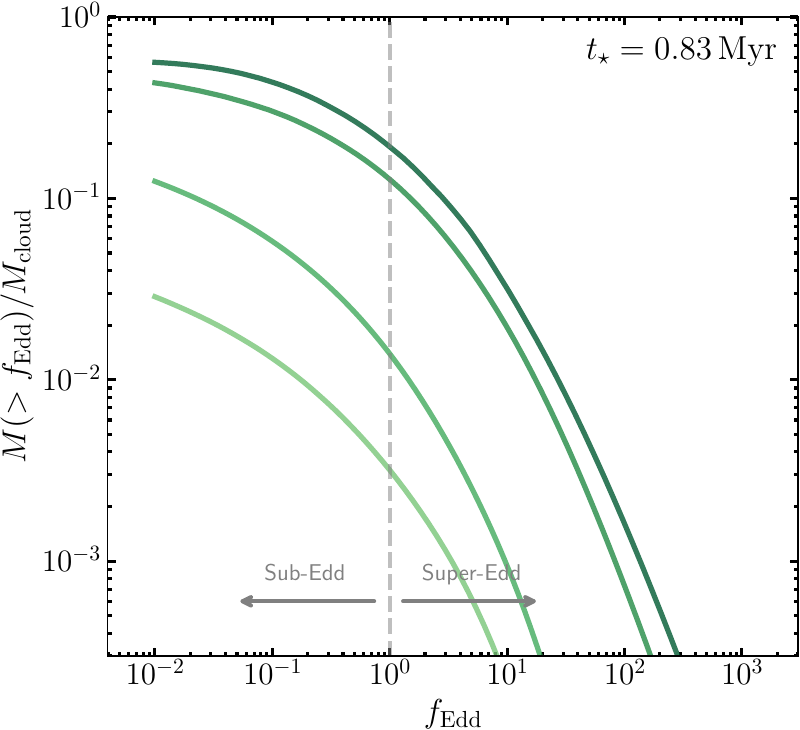}
  \caption{Distributions of the $\Lya$ Eddington ratio, $\fEdd$ (Equation~\ref{eq:fEdd}), shown as the cumulative fraction of $\Mcloud$ with $\fEdd$ greater than a given value at the time $t_\star = 0.83 \, \rm Myr$ where the dust-free case peaks in Figure~\ref{fig:fEddSnapshots}. The dashed gray line demarcates the $\fEdd = 1$ line. We can see that significant amounts of gas mass have high values of $\fEdd \gtrsim 10$, with densities $n \sim 10^2$--$10^4 \pcm$ (see Figure~\ref{fig:fEdd_vs_n}) suggesting that $\Lya$ radiation pressure can drive dynamically relevant winds.}
  \label{fig:fEdd_distributions}
  
\end{figure}

On the other hand, once the SFE has saturated, $M_{\fEdd>1}/\Mcloud$ peaks for all the dust cases, with considerable fractions of the remaining gas mass in the simulation ($M_{\rm gas}$) likely to be ejected were $\Lya$ present. For instance, for the dust-free limit, at $t_\star \sim 1 \, \rm Myr$, $M_{\rm gas}/\Mcloud \sim 40\%$ and $M_{\fEdd>1}/\Mcloud \sim 20\%$, suggesting that a substantial fraction of the leftover mass can be ejected by $\Lya$ radiation pressure. The peak values are indeed lower for the higher dust abundances, suggesting more modest differences would apply here than what is seen in the simulations without $\Lya$ feedback. 

While Figure~\ref{fig:fEddMainPlot}, and the preceding discussion quantifies the global fraction of super-Eddington mass, we can visually see from Figure~\ref{fig:fEddSnapshots} that there is a broad range of values of $\fEdd$. In Figure~\ref{fig:fEdd_distributions} we quantify these distributions at the time $t_\star = 0.83 \, \rm Myr$ (middle-left Figure~\ref{fig:fEddSnapshots}) when the dust-free case peaks, showing the cumulative amount of mass (normalized to $\Mcloud$) with Eddington ratio greater than a given $\fEdd$. The dashed gray line shows the trans-Eddington ($\fEdd=1$) line, and demarcates the fraction $M_{\fEdd>1}/\Mcloud$ at this time. We find that not only is a significant fraction of mass super-Eddington, there are also considerable amounts of mass with high $\fEdd$ values, e.g. $\approx 5\%$ of $\Mcloud$ or $\approx 5 \times 10^4 \, \Msun$ has $\fEdd >10$ for the dust-free case. This has implications on the dynamics of the outflows/winds ejected locally from the cluster-forming sites, with such high $\fEdd$ due to $\Lya$ radiation pressure likely implying large velocities---as has been previously suggested for radiation pressure driven-winds due to UV and dust-reprocessed IR \citep{Murray_2011,Thompson_2015, Menon_2023}. However, the effects are likely more dramatic for the $\Lya$ radiation pressure as the $\fEdd$ values obtained for the UV (IR) radiation pressure are typically $\lesssim 10$ ($\lesssim 1$), with the upper range limited to relatively low-density gas/sightlines \citep[see Figure 7 of][]{Menon_2025}, which contain relatively low amounts of mass. On the other hand, most of the mass of gas with high Eddington ratios ($\fEdd>10$) to $\Lya$ is likely coming from much denser gas with $n \sim 100$--$10^4 \pcm$. This gas is dense enough to be bright in recombination and forbidden emission lines, and therefore likely to have implications for observed line profiles. We will discuss these possible implications in Section~\ref{sec:discussion_winds}.

\subsection{Effectiveness of Lyman $\alpha$ radiation pressure}

\begin{figure*}
  \centering
  \includegraphics[width=\textwidth]{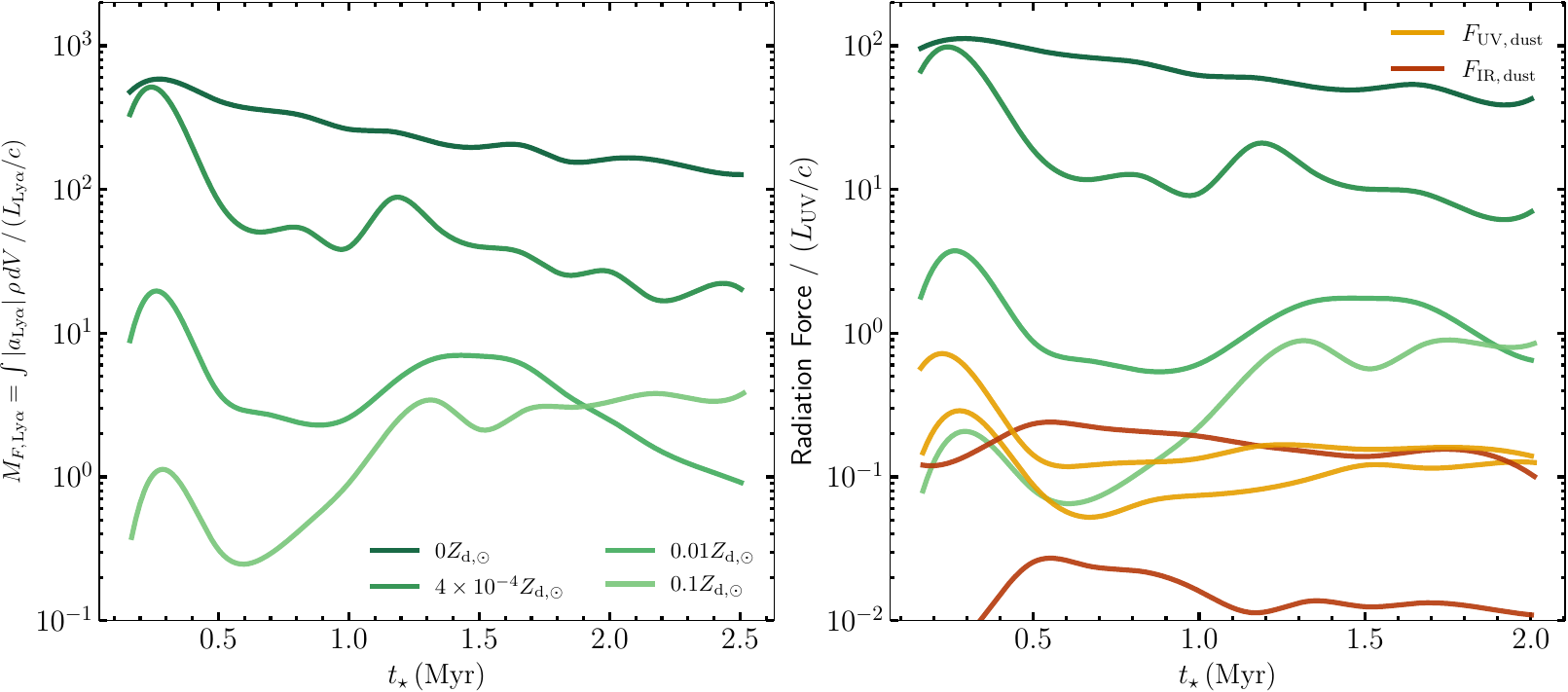}
  \caption{The $\Lya$ force multiplier (Equation~\ref{eq:Mf}; left) and the $\Lya$ and dust radiation pressure forces normalised to $L_{\rm UV}/c$ (right) as a function of the age of the stellar population ($t_\star$). We can see that $M_{\rm F}$ depends sensitively on the dust abundance, and decreases with time as the gas column in the cloud decreases (Figure~\ref{fig:fEddMainPlot}). However, at all dust abundances considered here, $\Lya$ radiation pressure dominates over the UV and IR radiation pressure forces.}
  \label{fig:Mf_vs_time}
\end{figure*}

In this section we quantify the momentum injection by $\Lya$ radiation pressure, and compare it to other relevant radiative feedback mechanisms that are included in the simulation on-the-fly (i.e. UV and IR radiation pressure on dust/gas). We compute the so-called $\Lya$ force multiplier $M_{\rm F}$ \citep{dijkstra_ly-driven_2008,kimm_impact_2018,smith_physics_2019,tomaselli_lyman-alpha_2021}, which represents the \textit{boost} in the total momentum imparted by $\Lya$ photons over the initial photon momentum (i.e. the single-scattering momentum $\sim L_{\Lya}/c$), akin to the force-multiplication factor used for metal absorption lines in stellar atmospheres \citep{Castor_1975} or the trapping factor ($\ftrap$) used for dust-reprocessed IR radiation \citep[e.g.,][]{Thompson_2005,Murray_2010,Menon_2022b}. Physically $M_{\rm F}$ is proportional to the ratio $t_{\rm trap}/t_{\rm light}$, where $t_{\rm trap}$ represents the time $\Lya$ photons remain trapped in a cloud due to resonant scattering, and $t_{\rm light}$ is the light-crossing timescale over the cloud, with classical analytical models predicting $M_{\rm F} \sim (a_{\rm V}\tau_{\rm cl})^{1/3}$ \citep{Adams_1975,Bonilha_1979,lao_resonant-line_2020}; the extremely high $\Lya$ optical depths in our clouds (see Figure~\ref{fig:Snapshots}) suggest $M_{\rm F} \sim 360 \, (\tau_{\rm cl}/10^{11})^{1/3} (T/10^4 \rm K)^{-1/6}$. That being said, continuum absorption by dust, velocity gradients, and anisotropic density structure may saturate $M_{\rm F}$ at lower values in the dusty turbulent media we consider in our simulations \citep{nebrin_lyman-_2025}.

We calculate an effective $M_{\rm F}$ for the whole cloud through the relation
\begin{equation}
  M_{\rm F} = \frac{\int |\aLya| \, \rho \, \text{d}V}{L_{\Lya}/c} \, ,
  \label{eq:Mf}
\end{equation}
i.e., comparing the integrated magnitude of radiative force to the single-scattering momentum of a $\Lya$ photon.

In the left panel of Figure~\ref{fig:Mf_vs_time} we show $M_{\rm F}$ as a function of time in our simulations for all the dust cases. We can see that $M_{\rm F}$ is relatively small ($\lesssim 10$) throughout the duration of the simulations for $\gtrsim 0.1 \, {\rm Z}_{\rm d,\sun}$, but can achieve values $\gtrsim 100$ for low dust abundances. The force multiplier also exhibits a clear decreasing trend with time across all dust cases except the $0.1 \, {\rm Z}_{\rm d,\sun}$, consistent with the reduction in the gas column density as the cloud is dispersed by radiative feedback and consumed by star formation (Figure~\ref{fig:fEddMainPlot}). The sensitivity of $M_{\rm F}$ to $Z_{\rm d}$ is striking---spanning nearly four orders of magnitude across the range of dust abundances considered---and underscores the critical role of dust in limiting the dynamical impact of Ly$\alpha$ radiation pressure through continuum absorption, which we discuss further in Section~\ref{sec:dust_discussion}.

We also compare the magnitude of $\Lya$ radiation pressure to that from other radiation pressure on dust included in the simulations, i.e., the direct UV and dust-reprocessed IR photons. For the conditions modeled in our simulations representative of star clusters at cosmic dawn ($\Sigma_{\rm cloud} \gtrsim 10^3 \, \rm \text{M}_{\sun} \, pc^{-2}$), the radiation pressure on dust is the dominant radiative feedback mechanisms \citep{Krumholz_Matzner_2009,Kim_2018,Menon_2023}, and therefore this comparison serves to inform us on the amount of momentum from pre-supernova feedback simulations without $\Lya$ are missing. In the right panel of Figure~\ref{fig:Mf_vs_time} we compare the total Ly$\alpha$ radiation pressure force to the UV ($F_{\rm UV,dust}$)\footnote{Here we include the cumulative force on dust from all three UV bands considered in the simulations: LyC ($h\nu > 13.6 \, \rm eV$), Lyman-Werner ($11.2 \, {\rm eV} < h\nu < 13.6 \, {\rm eV}$), and photoelectric/FUV ($6 \, {\rm eV} < h\nu < 11.2 \, {\rm eV}$).} and IR ($F_{\rm IR,dust}$) radiation pressure forces on dust, all normalized to the single-scattering momentum flux $L_{\rm UV}/c$. Ly$\alpha$ radiation pressure dominates over both the UV and IR forces across all dust abundances and at essentially all times. Even for the highest dust abundance case ($0.1 \, {\rm Z}_{\rm d,\sun}$), $\Lya$ radiation pressure is higher than UV and IR radiation pressure by a factor $\sim 10$ at $t_{*}> 1\, \rm Myr$, whereas for the dust-free limit the factor goes up to $\gtrsim 500$ at all times. The high factors are partly also driven by the fact that $F_{\rm UV,dust} \sim 0.1 L_{\rm UV}/c$, a factor of 10 lower than the total momentum available in the UV radiation field. This occurs due to the cancellation of radiative fluxes from multiple sources when the UV radiation is absorbed on scales smaller or comparable to the radius of the star cluster \citep[see][]{Menon_2023}. This also makes the IR force---which doesn't suffer from this as it diffuses through the IR-thick region---comparable to the UV force, as the dust IR optical depth approaches $\sim 0.1$ for $\Sigma_{\rm cloud} \gtrsim 10^3 \, \rm M_{\sun} \, pc^{-2}$ \citep{Menon_2022b}, facilitating a total force $\propto \tau_{\rm IR} L_{\rm UV}/c$. At lower dust abundances, however, the IR force falls well below the UV force, as the medium becomes increasingly optically thin to IR photons. Nevertheless, the dominance of Ly$\alpha$ radiation pressure over all of these mechanisms---even at the highest dust abundances considered---highlights that it represents a potentially critical and hitherto neglected feedback channel in dense star-forming systems, motivating its self-consistent inclusion in future RHD simulations of star cluster formation and the ISM \citep{nebrin_lyman-alpha_2026}.

\section{Semi-analytic models to quantify the effects of $\Lya$ radiation pressure}
\label{sec:semi-analytic}

\begin{figure*}
  \centering
  \includegraphics[width=\textwidth]{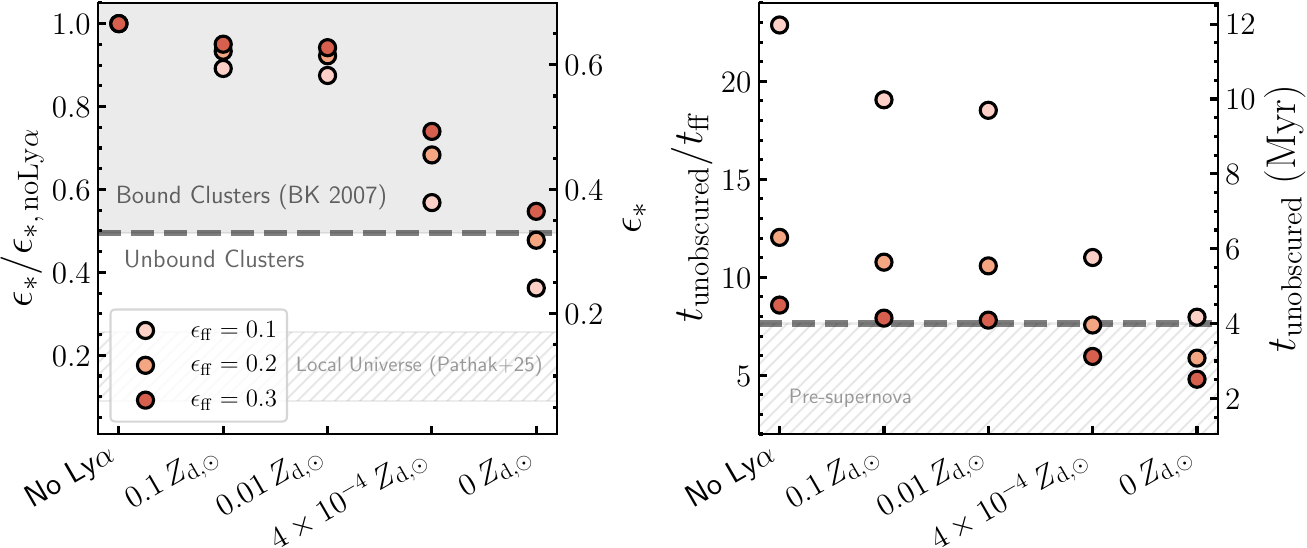}
  \caption{Plot quantifying the potential impact of $\Lya$ radiation pressure on stellar growth using a semi-analytic model(Appendix~\ref{sec:Appendix_TK16}) informed by insights from our calculations. \textbf{Left}: The integrated efficiency of star formation ($\epsilon_* = M_\star/M_{\rm cloud}$) for the different dust abundances \textit{relative} to the case without $\Lya$ radiation pressure (unfilled marker), with the parallel y-axis indicating the corresponding \textit{absolute} value appropriately scaled to the final stellar mass obtained in our simulations. Colors indicate different assumptions for $\epsilon_{\rm ff}$, the \textit{rate} at which stars form per free-fall time ($t_{\rm ff}$), with the shaded gray region indicating a rough threshold $\epsilon_*>0.33$ where bound cluster formation is likely dominant \citep[][BK2007]{Baumgardt_2007}; the hatched region shows the range of $\epsilon_*$ values inferred for \HII regions in local Universe galaxies in the PHANGS survey \citep{Pathak_2025}. \textbf{Right}: $t_{\rm unobscured}$, defined as the timescale for $>50\%$ of the surface area around the stellar population to be occupied by gas with $\Sigma \lesssim 10 \, \Msolpc$ which corresponds to $ A_{\rm V} \sim 1$ for ${\rm Z}_{\rm d,\sun}$. We see that while the inclusion of $\Lya$ radiation pressure noticeably decreases $\epsilon_*$, $\epsilon_*$ remains relatively high compared to less dense clouds in the local Universe, and may still go on to form highly bound clusters. On the other hand, the inclusion of $\Lya$ is able to eject gas and become unobscured more rapidly, suggesting implications for the attenuation of the cluster and ionizing photon escape.}
  \label{fig:SFE_TK16}
\end{figure*}

Our results demonstrate that the inclusion of $\Lya$ radiation pressure can render noticeable fractions of gas in star-cluster forming environments super-Eddington---up to $\sim 40\%$ (3\%) for dust-free ($0.1\,\text{Z}_{\sun}$) cases (Figure~\ref{fig:fEddMainPlot}). However, we also find that there is a time evolution in the super-Eddington fraction, with $\fEdd$ substantially lower at the early stages of the evolution when the bulk of the stellar mass is assembled. This complicates our attempts to quantify the potential impacts of $\Lya$ radiation pressure on the star formation efficiency, as one needs to examine the competition between gravity and radiation pressure simultaneously with the stellar assembly. This necessitates modeling $\Lya$ feedback on-the-fly in 3D RHD simulations, which remains beyond the capabilities of existing codes. Nevertheless, we can employ semi-analytic models of star formation, which have been tested on similar numerical simulations of radiation pressure-regulated star formation, to explore the relative impact of $\Lya$ radiation pressure on the star formation efficiency.

\subsection{Spherically symmetric + instantaneous ejection}
The simplest version of this considers the minimum fraction of gas ($\epsilon_*$ that needs to be assembled such that feedback is able to unbind the gas cloud \citep[e.g.,][]{Fall_2010,Grudic_2018}, given by
\begin{equation}
    \label{eq:SFEinst}
    \epsilon_* = \frac{\Sigma_{\rm tot}}{\Sigma_{\rm tot}+\Sigma_{\rm crit}} \, ,
\end{equation}
where $\Sigma_{\rm tot}$ is the total gas+stellar surface density, and 
\begin{equation}
    \Sigma_{\rm crit} = \frac{ \langle \dot{p}/M_\star \rangle}{4\pi G} \, ,
\end{equation}
is the critical surface density beyond which feedback at a given momentum injection rate $\langle \dot{p}/M_\star \rangle$ will be unable to regulate star formation. For $\Lya$ radiation pressure, 
\begin{equation}
    \langle \dot{p}/M_\star \rangle = \frac{M_{\rm F}L_{\Lya}}{cM_\star} = \frac{f_{\rm trap,\Lya}L_{\rm UV}}{cM_\star} \, ,
\end{equation}
where we define $f_{\rm trap,\Lya}$ as the momentum boost over the single scattering momentum $L_{\rm UV}/c$ as shown in the right panel of Figure~\ref{fig:Mf_vs_time}\footnote{We note that the mapping between $M_{\rm F,\Lya}$ and $f_{\rm trap,\Lya}$ depends on the the ratio of $L_{\Lya}/L_{\rm UV}$, which would depend on the spectrum of the source, pre-absorption or significant escape of LyC photons, and any additional contributions from collisionally excited $\Lya$ emission.}. This implies values of
\begin{equation}
    \label{eq:Sigmacrit_Lya}
    \Sigma_{\rm crit,\Lya} \sim 7000 \left( \frac{f_{\rm trap,\Lya}}{10}\right)\!\left(\frac{ L_{\rm UV}/M*}{2000 \rm L_{\sun} \Msun^{-1}} \right) \Msun \, \rm pc^{-2},
\end{equation}
where we have normalized to the UV light-to-mass ratio averaged over the first 3 Myr for a \texttt{BPASS} stellar evolution model, and a value of $f_{\rm trap,\Lya}$ appropriate for the $4\times 10^{-4} \, {\rm Z}_{\rm d,\sun}$ case (Figure~\ref{fig:Mf_vs_time}). Adopting the (time-averaged) values of $f_{\rm trap,\Lya}$ in Equation~(\ref{eq:SFEinst}), we obtain values of $\epsilon_*$ of 29\%, 27\%, 5\% and 1\% for the four values of ${\rm Z}_{\rm d,\sun}$ respectively---substantially lower than the $\sim 50\%$ efficiency this model predicts without $\Lya$. However, this model is overly simplified as it does not contain any timescale information, i.e., gas is assumed to be instantly ejected once the threshold $\epsilon_*$ is achieved and ignores the time to reverse the accretion flow ($\sim t_{\rm ff}$). Moreover this neglects the effects of turbulence, due to which considerable amounts of gas may be at higher $\Sigma > \Sigma_{\rm crit}$ than the average $\Sigma_{\rm tot}$.

\subsection{Turbulent density + gradual ejection}
To incorporate these effects, we use a slightly modified version of the \citet{Thompson_Krumholz_2016} model, described in Appendix~\ref{sec:Appendix_TK16}, which considers the competition between momentum injection by stellar feedback and gravity in turbulent self-gravitating clouds. This is similar to other models in the literature attempting to capture this competition \citep[e.g.,][]{Krumholz_Matzner_2009,Fall_2010}, with the distinction that \citet{Thompson_Krumholz_2016} takes into account the broad distribution of gas column densities seen by the stellar population, capturing the relative effects of feedback on high- and low-density sightlines---an important aspect for $\Lya$, as shown in Figure~\ref{fig:fEdd_vs_n}. This is done by assuming a lognormal shape for the mass/area weighted gas column density distribution seen by the point source representing the stellar population, with star formation (gas ejection) changing the mean of this distribution over time as gas is consumed (ejected). Similar to above, we use the net momentum injection per unit stellar mass for $\Lya$ radiation pressure based on the time-averaged value shown in the right panel of Figure~\ref{fig:Mf_vs_time}\footnote{These values are likely an overestimate, since the additional force from $\Lya$ and the associated gas ejection would lower gas columns and therefore the momentum it imparts on the gas. Moreover, this estimate does not consider the direction in which the $\Lya$ force acts, and assumes the entire momentum deposition acts in opposition to gravity.}, and adopt the initial cloud conditions identical to those considered in this work. The model needs two parameters: the turbulent Mach number $\mathcal{M}$, and the star formation efficiency per free-fall time $\epsilon_{\rm ff}$. For the former, we use the value drawn from our simulation ($\sim 20$); while we obtain a value of $\epsilon_{\rm ff} \sim 0.3$ in our simulations, we also explore lower values as the inclusion of $\Lya$ and other physics (e.g. magnetic fields) can lower $\epsilon_{\rm ff}$ \citep[e.g.,][]{Kim_2021}. 

In Figure~\ref{fig:SFE_TK16} we show the outcomes of the models for the different dust abundances, compared to the case without $\Lya$. To remove any dependencies on the parameter choices of our semi-analytic model, we show the integrated star formation efficiency $\epsilon_* = M_\star/\Mcloud$ \textit{relative} to the case without $\Lya$, and in the parallel axis, show the corresponding \textit{absolute} value, using the value of $\epsilon_*$ obtained in our RHD simulations as the value for the No $\Lya$ case. We can see that our model predicts modest changes to $\epsilon_*$ for $Z_{\rm d} \gtrsim 0.01 \, {\rm Z}_{\rm d,\sun}$---consistent with the relatively low levels of momentum injection ($\sim L_{\rm UV}/c$) for these cases. On the other hand, for lower dust abundances $\epsilon_*$ noticeably decreases, down to $\sim 20-40\%$ for the dust-free case, suggesting that $\Lya$ is an important regulator of star formation in dust-poor environments. That being said, we note that even in the dust-free limit, $\epsilon_*$ is much higher than those obtained in numerical simulations \citep[e.g.,][]{Kim_2018,Grudic_2022,Menon_2025} and in observations \citep{Chevance_23,Pathak_2025} of lower-density star-forming clouds typical of the local Universe ($\sim 10\%$); we show this band of values with hatches in Figure~\ref{fig:SFE_TK16}. We also show the minimum value of $\epsilon_* \sim 0.33$ for which there is considerable amounts of bound star formation in the $N$-body calculations of \citet{Baumgardt_2007} to indicate that $\Lya$ is unlikely to preclude bound cluster formation, except possibly in dust-free environments. This suggests that $\Lya$ radiation pressure is unlikely to alter the picture of efficient star formation in dense cluster-forming environments in galaxies at cosmic dawn.

On the other hand, the inclusion of $\Lya$ radiation pressure is able to eject a larger fraction of gas on faster timescales. In the right panel, we show $t_{\rm unobscured}$, defined as the timescale for $>50\%$ of the surface area around the stellar population to be occupied by gas with $\Sigma \lesssim 10 \, \Msolpc$. This threshold $\Sigma$ was chosen as it corresponds to $ A_{\rm V} \sim 1$ for ${\rm Z}_{\rm d,\sun}$, and is simply meant to represent low-column density sightlines through which UV and LyC photons could potentially escape. We can see that compared to the case without $\Lya$, $t_{\rm unobscured}$ is shorter for all cases, and significantly so for the dust-poor and dust-free cases. This would have implications for the attenuation of radiation from the stellar population, and the ability of this radiation to affect the ISM of their host galaxies for longer periods before supernovae feedback becomes relevant ($\gtrsim 4 \, \rm Myr$; shaded region). Moreover, clearing gas sightlines and subsequent ionization of these channels are critical for Lyman continuum photon escape---$\Lya$ driven-winds could facilitate this on shorter timescales, further permitting LyC escape from cluster-forming sites before the source of these photons (i.e.\ massive stars) die.

\section{Discussion}
\label{sec:discussion}

Our results suggest a consistent picture in which $\Lya$ radiation pressure is a critical source of early stellar feedback in dense star cluster-forming regions at dust abundances typical of cosmic dawn ($\lesssim 0.1 \rm\,\text{Z}_{\sun}$), dominating over UV and IR radiation pressure on dust/gas. However, at the same time, the \textit{effective} impact of $\Lya$ radiation pressure depends on its competition with gravitational forces ($\fEdd$), for which we find a diverse range of outcomes depending on the density of the gas: lower density ($n \lesssim 10^3 \, \pcm$) gas has $\fEdd>1$, whereas $\fEdd<1$ for the high-density ($n > 10^4 \, \pcm$) filaments through which most of the stellar growth occurs. Therefore our results suggest that the star formation efficiency is likely to remain high in the dense star clusters forming at high-$z$. On the other hand, a significant fraction of the volume-filling gas can be ejected by $\Lya$ radiation pressure, suggesting implications for the visibility of light from the cluster, ionizing photon escape, and launching outflows from the galaxy.

\subsection{Star formation in galaxies at cosmic dawn}
\label{sec:discussion_starformation}

Recent works have alluded to the inability of pre-supernova stellar feedback to regulate the efficiency of star formation in dense star-forming clouds \citep[e.g.,][]{Grudic_2018,Menon_2023,Menon_2025} to suggest that galaxies form stars more efficiently in the (relatively) massive galaxies observable with the \textit{JWST} at cosmic dawn \citep{Dekel_2023,Somerville_2025}---one of several suggestions that may alleviate the tension in the empirical excess of UV-bright galaxies over pre-launch predictions \citep[e.g.,][]{Finkelstein_2023}. However, the large amounts of momentum injection expected from $\Lya$ radiation pressure \citep{nebrin_lyman-_2025} were not considered as a feedback mechanism in these studies, which places into question the foundation of efficient star formation in dense GMCs that these solutions relied on. Indeed, semi-analytic models \citep{Ferrara_2025,manzoni_lyman-_2025} and simulations with subgrid models calibrated on idealized calculations \citep{kimm_impact_2018} find that it can significantly suppress the formation of bound clusters and reduced the integrated stellar mass formed in clusters and their host galaxies---except for the most extreme range of conditions ($\Sigma_* \gtrsim 10^5 \, \Msolpc$). 

Our results, on the other hand, indicate that even for $\Sigma_* \sim 10^3 \, \Msolpc$, typical for star clusters in galaxies at $z \gtrsim 2$ \citep{adelaide_first_2026}, $\Lya$ is unlikely to prevent high gas-to-star conversion ($\epsilon_* \gtrsim 50\%$). Counter-intuitively, this is not because $\Lya$ does not deposit significant amounts of momentum to the gas\footnote{This is a subtlety that is important to consider: radiation may be effectively trapped and generate significant amounts of momentum, but that momentum does not couple to the surrounding ISM unless $\fEdd>1$, as it just gets accreted back onto the stellar sources. Therefore applying the momentum in models/simulations without considering the feedback--gravity competition would overestimate the impact of feedback \citep{Krumholz_2018,Menon_2022b}.}---momentum injection can be up to $\sim 100 \, L_{\rm UV}/c$ at low dust abundances, dominating over UV/IR radiation pressure at effectively \textit{all} dust abundances. Rather, it is because $\Lya$ is unable to compete with the deep gravitational potential wells of  gas in the turbulent medium---an effect that is not captured by idealized models assuming a uniform distribution of gas at the mean density (Section~\ref{sec:semi-analytic}). Using a semi-analytic model that attempts to (approximately) capture these effects of a turbulent medium (Section~\ref{sec:semi-analytic}), and using $\Lya$ momentum injection rates calculated from our simulations, we find that the inclusion of $\Lya$ can only reduce the stellar mass formed in $\Sigma \sim 10^3 \, \Msolpc$ GMCs by $\sim 10\%$ for dust abundances of $ \gtrsim 0.01 \, {\rm Z}_{\rm d,\sun}$ ensuring $\epsilon_* \gtrsim 60\%$. Even for very dust poor ($\sim 10^{-4} \, {\rm Z}_{\rm d,\sun}$) and dust-free environments, $ \epsilon_* \gtrsim 30\%$, much larger than the $\epsilon_* \sim 10\%$ values seen in star-forming regions in the local Universe \citep{Pathak_2025}.

Although $\Lya$ (and other feedback mechanisms) remains unlikely to regulate the efficiency of star formation in dense cluster-forming sites, we expect it to be more effective in lower density star clusters in these galaxies and render them unbound. This is because to zeroth order, the gravitational acceleration is $\propto \Sigma$, whereas the acceleration imposed by $\Lya$\footnote{Technically this is $\dot{p}/M_*$, the momentum injection rate per unit stellar mass formed, and is the relevant quantity for momentum-feedback regulated star formation models \citep[e.g.,][]{Fall_2010,Thompson_Krumholz_2016,Grudic_2018}} is $\propto M_{\rm F} \propto \tau_{\rm \Lya}^{1/3} \propto \Sigma^{1/3}$, i.e., the competition between $\Lya$ radiation pressure and gravity skews more favorably to feedback at lower $\Sigma$ regions\footnote{This assumes that the $\tau_{\Lya}^{1/3}$ scaling relevant to optically-thick conditions apply to these regions.}. These considerations might be particularly relevant to galaxies in lower-mass halos ($\lesssim 10^8 \, \Msun$ at $z\sim 10$) which would have lower average baryonic surface densities ($\propto M_{\rm h}^{1/3}$) and metallicities/dust abundances, where $\Lya$ might primarily produce unbound star clusters, which likely go on to evolve to become ultra-faint dwarf galaxies in the present-day Universe \citep{Ricotti_2016}. It is critical that numerical simulations that aim to capture their formation and evolution to present-day relics include $\Lya$ feedback, as also pointed out by \citet{nebrin_lyman-_2025}. In a follow-up paper, we will extend the analysis of this paper to a broader range of GMC conditions and metallicities to quantify the effectiveness of $\Lya$ in different environments across cosmic time. 

Another aspect to keep in mind is that if the IMF were top-heavy in the high-$z$ Universe, e.g. due to lower gas-phase metallicities \citep{Sharda_2022,Chon_2023} and/or PopIII contributions \citep{Vendetti_2023,Zier_2025,vanVeenen_2026}, the momentum injection from $\Lya$ radiation pressure, and its competition with gravity could play out differently. For instance, \citet{Menon_2024} showed that a (very) top-heavy IMF induced light-to-mass ratio could lower the  stellar mass formed by up to a factor of $\sim 2$ for clusters with $\Sigma_* \sim 10^3 \, \Msolpc$ by UV radiation pressure on dust/gas. This would decrease even more significantly were $\Lya$ radiation pressure considered, especially if a top-heavy IMF and low gas-phase metallicities are linked. The extreme limit of these considerations are realized in the formation of Population III star clusters in metal-free mini-halos \citep{Klessen_2023}. Several radiation-hydrodynamic simulations resolving the AU-scale interaction of radiation feedback from PopIII stars have found that \HII regions produced by these stars are trapped for relatively long periods by the high gas columns and densities realized in their protostellar disks \citep{Jaura_2022,Sharda_Menon_2025,Chen_2026}, permitting high accretion rates and rapid stellar growth, with possible implications for the formation of extremely massive and/or supermassive stars and black holes with Ly$\alpha$ winds \citep{Smith_2016, Smith_2017, Smith_2017_DCBH}. Potentially, the inclusion of $\Lya$ radiation pressure could facilitate the \HII regions to break out from the compact phase in the polar directions \citep{Mckee_2008} and photoevaporate the ambient gas that feeds the disk/star system, effectively limiting the final mass of the star/cluster \citep{hosokawa_protostellar_2011}\footnote{However, the large \HI column densities realized in PopIII disks, and the deep potential well associated with might still be too high even for $\Lya$ radiation pressure in dust-free media \citep{nebrin_lyman-_2025}.}, and even potentially, the PopIII IMF. 

\subsection{Ionizing photon escape and wind launching}
\label{sec:discussion_winds}

While we find that most of the high-density gas---which carries substantial fractions of the total mass---is sub-Eddington to $\Lya$ radiation pressure, most of the volume is in lower density gas which is super-Eddington (Figure~\ref{fig:fEdd_vs_n}). This can be confirmed visually in Figure~\ref{fig:fEdd_distributions}, where a large fraction of the solid angles around the (central) stellar population has $\fEdd>1$, suggesting that they would be ejected by $\Lya$ radiation pressure. Moreover, our semi-analytical star formation model indicates that progressively larger fractions of the gas mass is ejected on shorter timescales as $\Lya$ becomes more effective (Figure~\ref{fig:SFE_TK16}). Such rapid, efficient dispersal of dense gas around dense cluster-forming regions likely has implications for the escape of UV photons from the stellar populations and the launching of outflows by the starburst. 

For instance, clearing the dense gas surrounding young stellar populations remains the most prominent bottleneck to the escape of LyC photons from galaxies typical of the high-$z$ Universe \citep{Paardekooper_2015,Trebitsch_2017}. The rapid drop of the LyC emission rate of young stellar populations as massive stars die in supernovae explosions ($\gtrsim 3 \, \rm Myr$) make models that allude to pre-supernova feedback in clearing sightlines for LyC escape attractive for facilitating the growth of ionized bubbles \citep{Carr_2024,Jaskot_2025}. Clearing dusty gas in the centers of galaxies has also been invoked to explain the extremely bright and blue spectrum of galaxies with minimal evidence of dust attenuation \citep{Ferrara_2023a,Ferrara_2023b,Marques-Chaves_2026}. On these fronts, numerical simulations of star cluster-forming clumps indicate that LyC and UV radiative feedback may already evacuate sightlines on $\sim 50 \, \rm pc$ scales for UV/LyC escape on $\lesssim 3 \, \rm Myr$ timescales through LyC and FUV radiation pressure on dust/gas after star formation saturates \citep{Menon_2025}, which may subsequently even launch outflows in dwarf galaxies with weak potential wells \citep{Thompson_2015,Komarova21a,Amorin_2024}. The highly elevated momentum (factors $>10$) injection rates over UV radiation pressure offered by $\Lya$ photons may substantially strengthen this picture, especially in the high-$z$ Universe with its lower dust abundances and halo potential wells. The high $\fEdd$ ratios ($>10$) realized in dust-poor environments (Figure~\ref{fig:fEdd_distributions}) even at relatively high densities ($n \sim 10^3 \pcm$; Figure~\ref{fig:fEdd_vs_n}) may also drive gas to large velocities exceeding the escape velocity \citep{Thompson_2015}, potentially contributing to the broad wings observed in emission line profiles of  LyC-leaking galaxies \citep{Komarova21a,Mainali_2024,Carr_2024}. We note, however, that $\Lya$ acceleration relies on maintaining high optical depths, and the expansion of gas in the outflow (lower $N_{\HI}$) and the lower cross-section by Doppler shifts \citep{dijkstra_ly-driven_2008} would limit its contribution to wind launching to the vicinity of the starburst \citep{Smith_2017}; nevertheless, the single scattering radiation pressure may continue to accelerate the gas beyond this point \citep{Zhang_2017}. There is scope to explore the acceleration of galactic winds through the combination of UV and $\Lya$ radiation pressure in future studies \citep{thompson_theory_2024}.

Even if $\Lya$+UV radiation pressure  is unable to directly launch outflows on galactic scales, especially in higher mass halos \citep{Smith_2017}, the pre-clearing of gas surrounding dense star clusters it facilitates would likely make supernova feedback more effective as explosions go off in lower-density gas \citep{Hennebelle_2014,Walch_2015,Kim_2015}. The lower densities reduce radiative cooling losses implying both higher momentum and energy injection rates \citep{Kim_2015}, which may lead to larger outflow velocities and hot gas ($T>10^6 \, \rm K$) fractions. The counterpoint is that more effective pre-supernova feedback may reduce the clustering of massive stars \citep{Smith_2021}, and the additional enhancement in momentum injection and the maintenance of hot bubbles offered by the overlap of the individual SNe bubbles \citep{Sharma_2014,Gentry_2017,Fielding_2017,ElBadry_2019}. Indeed, \citet{kimm_impact_2018} found in their simulations that the inclusion of a subgrid model for $\Lya$ radiation pressure decreased local star formation efficiencies, and the clustering of supernovae, producing lower mass outflow rates and hot gas fractions. However, as discussed below, injecting the momentum without resolving the density substructure will likely overestimate its effects on regulating local star formation, therefore underestimating the degree of the clustering of star formation. Applying the insights from this study would imply that the local star formation efficiencies and degree of clustering would still be high with the inclusion of $\Lya$. The combination of high clustering with rapid gas dispersal locally would imply highly efficient winds that entrain significant fractions of the supernova energy \citep{Fielding_2018}, which may further contribute to highly bursty star formation histories \citep[e.g.,][]{McClymont_2025_bursts}.

\subsection{Nuances imposed by clumpy gas distributions}
\label{sec:discussion_turbulence}

Turbulence driven by a combination of sources in the ISM induces clumpy, anisotropic gas distributions and disordered velocity fields \citep[e.g.,][]{Federrath_2008} that influence the impact of radiation pressure on dust/gas in \HII regions \citep{Raskutti_2016,Kim_2017,Menon_2022b,Menon_2023}. For the case of $\Lya$ radiation pressure, the impact of turbulence is borne out in several ways. For instance, turbulent density fluctuations can produce channels with low $\tau_{\Lya}$ through which photons can escape without significant momentum  \citep{kakiichi_radiation_2021}. \citet{nebrin_lyman-_2025} estimate that the $\Lya$ force multiplier $M_{\rm F,\Lya}$ is suppressed by a factor $\sim 10$ for gas with turbulent Mach numbers of $\mathcal{M} \sim 10$ with a scaling $M_{\rm F,\Lya} \propto \mathcal{M}^{-8/9}$. Preliminary comparisons of the $M_{\rm F,\Lya}$ we deduce in our calculations with analytic uniform density sphere/slab solutions suggest similar levels of suppression, although we defer a systematic comparison of models to a follow-up paper. Suppression of $M_{\rm F,\Lya}$ by factors $\sim$few may also be introduced by velocity fluctuations in the turbulent flow or in outflowing gas  \citep{nebrin_lyman-_2025}, although only for high relative velocities. 

Our calculations on the other hand highlight yet another effect introduced by turbulent density distributions: namely, the arrangement of gas into filamentary structures, with gas densities much higher (and lower) than the average \citep{Federrath_2008,Federrath_klessen_2012}. This has implications on the competition between $\Lya$ radiation pressure and gravity: i.e. even for a given (average) value of $M_{\rm F,\Lya}$ and $\Lya$ momentum injection rates, higher column density filamentary structures have lower values of $\fEdd$, with $\fEdd<1$ for sufficiently high $\Sigma > \Sigma_{\rm crit}$ where $\Sigma_{\rm crit}$ is given by Equation~(\ref{eq:Sigmacrit_Lya}) for $\Lya$. This implies that even if the \textit{average} column of gas is  $\lesssim \Sigma_{\rm crit}$---which is the case for the clouds modeled in this work---a noticeable fraction of gas mass that are at $> \Sigma_{\rm crit}$ remain resistant to $\Lya$ radiation pressure, necessitating larger efficiencies than implied by assuming a uniform density. In fact, this nuance is true for all forms of radiation pressure \citep{Raskutti_2017,Menon_2023}. A corollary of this point is that if numerical simulations do not resolve the density substructure in \HII regions, where the competition between radiation pressure and gravity play out, they may overestimate the impact of feedback \textit{even if they model the radiative transfer with a perfect method}. Moreover, there is the added concern raised by \citet{Krumholz_2018} that it matters at what scale the radiation pressure--gravity competition plays out, and applying momentum at larger scales necessitated by the numerical resolution may further bias towards feedback. 

The aforementioned points likely influence the strong effects on bound cluster formation found in the galaxy-scale simulations with a subgrid $\Lya$ feedback model in \citet{kimm_impact_2018}, where they impart a force of $\sim M_{\rm F}L_{\Lya}/c$ uniformly around star particles based on the local gas column of the cell. The values of $M_{\rm F}$ they use are likely overestimated as it does not consider suppression by turbulence and velocity gradients \citep{nebrin_lyman-_2025}, and the lack of resolved substructure precludes the survival of high-density filaments that continue to feed star formation\footnote{We note, however that their MC calculations likely underestimated the values of $M_{\rm F}$ by a factor $\sim 3$ due to a numerical bug \citep{nebrin_lyman-_2025}, which would act to bring it closer to values considering these physics.}. These challenges motivate the development of subgrid models of $\Lya$ feedback calibrated to high-resolution turbulent clouds in the future.

\subsection{Critical role of dust opacity}
\label{sec:dust_discussion}

Our findings confirm that the destruction of $\Lya$ photons by dust continuum absorption is critical to the amount of momentum it imparts to the gas and its impacts on star formation and winds. For instance, the effects of $\Lya$ radiation pressure on our clouds are relatively mild for $Z_{\rm d} \gtrsim 0.01 \, {\rm Z}_{\rm d,\sun}$, indicating that if dust abundances in galaxies at cosmic dawn are similar to those found in the local Universe, the inclusion of $\Lya$ is unlikely to dramatically alter the outcomes of feedback-regulated star formation at these epochs. This suggests that cluster-forming sites in the (relatively) massive galaxies observable with the \textit{JWST} at the EoR ($M_\star \gtrsim 10^8 \, \Msun$, which have gas-phase metallicities $\gtrsim 0.1 \, Z_{\rm sun}$ \citep[e.g.,][]{Stanton_2026} are only likely to be weakly affected by $\Lya$, with strong feedback limited to low-mass galaxies which indeed dominate the population of galaxies at these epochs, but are not easily observed with the \textit{JWST}.

However, as alluded to in Section~\ref{sec:methods}, there is possible empirical evidence that the dust-to-metal ratio drops in low-metallicity galaxies below $0.1\,\text{Z}_{\sun}$ \citep{remy-ruyer:2014,deVis_2019}, indicating a significantly lower value of $Z_{\rm d}$ even at the relatively high gas-phase metallicities in \textit{JWST}-observed galaxies---indeed, the case with $Z_{\rm d} \sim 4 \times 10^{-4} \, {\rm Z}_{\rm d,\sun}$ corresponds to the dust abundance assuming the empirical scaling of \citet{remy-ruyer:2014} at a gas phase metallicity of $0.01\,\text{Z}_{\sun}$. We note that the dust-to-metal ratio may be even lower than this empirical scaling at high redshift ($z \gtrsim 10$) where dust production may be dominated by SNe as opposed to Asymptotic Giant Branch (AGB) stars, which may be prone to lower effective dust yields due to destruction in the reverse shock in a supernova explosion \citep[][and references therein]{Schneider_2024}.

Moreover, we stress that the critical quantity is the dust optical depth at (rest-frame) $\Lya$ wavelengths, not uniquely the dust abundance\footnote{Formally, the outcome depends on the ratio of optical depths of the gas cloud $\epsilon = \tau_{\rm dust}/\tau_{\Lya}$, where $\tau_{\rm dust} = \Sigma_{\HI} \kappa_{\rm dust,\Lya} {\rm Z}_{\rm d,\sun}$, with the role of dust effectively to saturate the value of $M_{\rm F}$ at high $\tau_{\rm Lya}$ to a value $\sim 3(a_{\rm V}/\epsilon)^{1/4}$ \citep{Neufeld_1991,Hansen_2006,tomaselli_lyman-alpha_2021,nebrin_lyman-_2025}.}. It is possible that even for a given $Z_{\rm d}$, the effective dust opacity could be lower in the early Universe due to distinct dust grain size distributions and/or compositions produced by interaction of dust production, growth, and destruction mechanisms in the ISM. For instance, some theoretical \citep{Hirashita_2019} and numerical simulations with on-the-fly dust modeling \citep{Narayanan_2025} have suggested that the grain size distribution peaks at larger sizes at $z \gtrsim 10$, implying UV dust extinctions can be reduced by up to an order of magnitude. The combination of lower dust-to-metal ratios and UV dust opacities might significantly elevate the impact of $\Lya$ photons even in relatively metal-rich systems, further motivating an understanding of the nature of dust at these epochs.

\subsection{Caveat of post-processing}

The primary caveat of our study is that we are estimating the (potential) dynamical impact of $\Lya$ radiation pressure for instantaneous gas snapshots in post-processing rather than including the physics self-consistently on-the-fly. While our analysis is highly informative on its competition with gravity, especially when interpreted using semi-analytic models of star formation (Section~\ref{sec:semi-analytic}), they cannot capture the nonlinear effects that may be present were $\Lya$ feedback included on-the-fly. Particularly, our findings reveal that the effectiveness of $\Lya$ feedback against gravity is sensitive to the fraction of gas mass that is concentrated in dense filaments with $\fEdd <1$ (Figure~\ref{fig:fEddMainPlot}), as opposed to the lower-density volume-filling phase with $\fEdd >1$. However, our post-processing approach misses the effects of $\Lya$ on this density distribution itself.

For instance, $\Lya$ ejecting lower-density gas and carving low-density channels would make it easier for subsequent gas ejection through thermal-pressure gradients of photoheated gas, and additional momentum injection on denser neutral filaments through the ``rocket effect'', i.e. by the back-reaction induced by the escape of photoevaporated gas (``champagne flow'') through these channels \citep{Whitworth_1979,Krumholz_2006,Kim_2018,Menon_2020,kakiichi_radiation_2021}. Moreover, ionizing photons can continually maintain photoevaporative flows off the dense neutral gas into the ionized phase if it is being evacuated \citep{bertoldi_photoevaporation_1989,bertoldi_photoevaporation_1990}, thereby lowering the mass fraction in the dense $\fEdd<1$ filaments. Similar effects are expected from other physics missing in our simulations that are likely to affect mass exchange between dense and diffuse phases: such as line-driven stellar winds\footnote{Earlier work has shown that momentum injection from stellar winds remain limited due to efficient cooling in turbulent mixing layers  \citep{Lancaster_2021b,Lancaster_2021c}, though the robustness of this result is sensitive to additional physics \citep{Lancaster_2024,Lancaster_2025,Rodriguez_2026}.}, magnetic fields, and ultimately, numerical resolution in the mixing layers. These points suggest that our post-processing predictions are overestimating the efficiencies of star formation. On the other hand, an expanding flow would also make the source distribution of $\Lya$ more diffuse and induce doppler shifts due to velocity gradients/outflows \citep{dijkstra_ly-driven_2008,Smith_2017}, all of which would act to lower the effectiveness of $\Lya$. 

These nonlinear interactions between the physics suggest that it remains to be seen how $\Lya$ would impact the assembly of star clusters. We therefore urge the community to treat these calculations as providing insight on the relative effects of $\Lya$, and as motivation to drive progress in modeling $\Lya$ radiation transport on-the-fly in tandem with these feedback mechanisms \citep[see][for recent progress]{nebrin_lyman-alpha_2026}.

\subsection{Numerical convergence}

Calculations using MC estimators are notoriously prone to slow ($\propto 1/\sqrt{N}$) convergence \citep{Camps_2018,Krieger_2023,Krieger_2024} especially for internal quantities such as the radiative acceleration and force multipliers \citep[see][for $\Lya$ MCRT]{Kasiri_2026}. In Appendix~\ref{sec:Appendix_numerical}, we conducted a convergence study with respect to the number of photons and found that there is a lack of convergence at the factor $\sim 2$ level, although in the direction of reinforcing our conclusions. Nevertheless, this suggests that the \textit{absolute} values of $\fEdd$ quantified with our cell-level estimators are potentially prone to MC noise and would need to be interpreted with caution. An alternative is to define a \textit{global} estimator for $\fEdd$ that considers the volume integrated force competition between $\Lya$ and gravity over the entire cloud, which will be robust to the effects of MC noise---we explore such an estimator in Section~\ref{sec:Appendix_fEdd}. However, as emphasized above, the broad density distributions achieved in the gas implies a diversity of outcomes of the gravity--feedback competition that can locally regulate collapse and facilitate gas ejection. The global estimator averages over this distribution with a bias to the bulk of the mass, missing the nuances imposed by the density distribution. Therefore, we prefer the locally-defined estimator for $\fEdd$ (i.e. Equation~\ref{eq:fEdd}) and systematically evaluate the impact of MC noise on our results.

To quantify the extent of the impact of MC noise, we conduct an experiment in Appendix~\ref{sec:Appendix_MCNoise} where we progressively increase the signal-to-noise of the $
\Lya$ radiation field by coarsening the spatial resolution of the domain, while maintaining the photon number to the fiducial value ($10^6$). We find that for cases where $\fEdd>1$ for noticeable fractions of the mass---i.e. at low dust abundances ($Z_{\rm d} \lesssim 0.01 Z_{\rm d,\sun}$)---the impact of MC noise is relatively mild (order $\sim$ unity) and the absolute fractions of super-Eddington gas (e.g. in Figure~\ref{fig:fEddMainPlot}) remain relatively unchanged. On the other hand, in dusty environments ($Z_{\rm d} \gtrsim 0.01 Z_{\rm d,\sun}$) $\fEdd$ is further lowered at higher signal-to-noise (by factors up to $\sim 5$). Taken together, this implies that our conclusion that in dust-poor environments $\Lya$ radiation pressure affects star cluster assembly is robust to MC noise and that it plays a negligible role in dusty environments is further reinforced.

\section{Conclusions}
\label{sec:conclusions}

We quantify the dynamical importance of $\Lya$ radiation pressure in dense star cluster-forming environments to interpret its potential to regulate star formation and eject gas in galaxies at cosmic dawn. We post-process snapshots of radiation-hydrodynamic simulations of the collapse of turbulent GMCs forming systems with properties resembling observed lensed star clusters \citep[$M_\star \gtrsim10^6 \, \rm M_\sun$, $\Sigma_* \gtrsim 10^3 \, \rm M_\sun \, pc^{-2}$][]{Adamo_2024,Claeyssens_2026} with the Monte Carlo Ly$\alpha$ radiative transfer code \textsc{colt}. We span a range of dust abundances, $Z_\mathrm{d} = \{0, 4\times10^{-4}, 0.01, 0.1\} \, Z_\mathrm{d,\sun}$, representative of the parameter space relevant to galaxies observed with the \textit{JWST} at the Epoch of Reionization. We quantify the competition between $\Lya$ radiation pressure and gravity, its dependence on gas properties and dust abundance, the momentum injected by $\Lya$ photons, and its relative importance to other radiative feedback mechanisms. We incorporate these insights into semi-analytic star formation models to infer the potential effects of $\Lya$ were it included on-the-fly. Our key findings are:

\begin{enumerate}[leftmargin=*]

\item \textbf{Ly$\boldsymbol{\alpha}$ radiation pressure is unlikely to preclude efficient star formation}: The fraction of the initial cloud mass that is super-Eddington ($\fEdd>1$) to $\Lya$ radiation pressure remains low at all times, $\lesssim 3 \%$ ($\lesssim 35\%$) for $0.1 \, {\rm Z}_{\rm d,\sun}$ (dust-free limit) respectively. Especially at early times when stellar mass is being assembled. This is because the densest gas that dominates stellar mass assembly ($n \gtrsim 10^4 \, \mathrm{cm}^{-3}$) has $\fEdd <1$ for all dust cases, with $\fEdd>1$ predominantly for the bulk of the volume-filling gas ($n \lesssim 10^3 \pcm$). Our semi-analytic model incorporating these insights find that the gas-to-star conversion efficiency $\epsilon_*$ would remain $\gtrsim 50\%$ with the inclusion of $\Lya$ in environments with non-zero dust abundances. Even in the dust-free limit $\epsilon_* \gtrsim 25\%$, much higher than values observed for typical clouds in the local Universe ($\sim 10\%$).

\item \textbf{Ly$\boldsymbol{\alpha}$ radiation pressure can drive dynamically significant winds once star formation saturates}: Once the SFE has saturated, $M_{f_\mathrm{Edd}>1}/M_\mathrm{cloud}$ peaks for all dust cases, with the super-Eddington mass fractions comparable to the total remaining gas mass, suggesting that Ly$\alpha$ can eject a significant fraction of the residual gas. Furthermore, $\gtrsim 5\% M_\mathrm{cloud} \sim 5 \times 10^4 \, \rm M_{\sun}$ has $f_\mathrm{Edd} > 10$ in the dust-poor environments, implying that a substantial amount of gas is susceptible to acceleration to large velocities. Our semi-analytic model indicates that $\Lya$-induced gas ejection can remove obscuration around the stellar sources on timescales quicker by up to factors $\sim 2$ to values $\lesssim 4 \, \rm Myr$, with implications for ionizing photon escape and the attenuation of radiation from the stellar population.

\item \textbf{Ly$\boldsymbol{\alpha}$ radiation pressure dominates over UV and IR radiation pressure at all dust abundances}: The Ly$\alpha$ force multiplier $M_\mathrm{F}$ is highly sensitive to the dust abundance, spanning nearly four orders of magnitude across our parameter space, from $M_\mathrm{F} \lesssim 10$ for $Z_\mathrm{d} \gtrsim 0.1 \, Z_\mathrm{d,\sun}$ to $M_\mathrm{F} \gtrsim 100$ at low dust abundances. Nevertheless, even at the highest dust abundance considered, Ly$\alpha$ radiation pressure exceeds the UV and IR radiation pressure forces by a factor of $\sim 10$, rising to $\gtrsim 500$ in the dust-free limit. This highlights Ly$\alpha$ radiation pressure as the dominant radiative feedback mechanism in dense, low-metallicity star-forming systems.

\item \textbf{Accurate dust opacities at Ly$\boldsymbol{\alpha}$ wavelengths are essential for robust feedback predictions}: Our results confirm that continuum absorption of Ly$\alpha$ photons by dust is a critical factor in regulating its dynamical impact, with dust-poor environments ($Z_\mathrm{d} < 0.01 \, \rm Z_\mathrm{d,\sun}$) highly susceptible to its effects. Given empirical evidence for a super-linear relation between metallicity and dust-to-gas ratio at $Z \lesssim 0.1 \, \rm Z_\sun$ \citep{remy-ruyer:2014}, and the possibility of distinct grain size distributions in the early Universe that could further reduce the dust opacity at Ly$\alpha$ wavelengths, the effective dust opacity may be considerably lower than a simple linear scaling with metallicity would suggest, underscoring the importance of careful dust modeling when assessing the role of Ly$\alpha$ feedback at cosmic dawn. In addition to making $\Lya$ feedback more effective, lower dust opacities would also boost the visibility of high-$z$ $\Lya$ emitting galaxies.
\end{enumerate}

We also discuss implications on star formation (Section~\ref{sec:discussion_starformation}), photon escape and winds (Section~\ref{sec:discussion_winds}), nuances imposed by clumpy gas distributions (Section~\ref{sec:discussion_turbulence})  and our assumptions on the dust opacity (Section~\ref{sec:dust_discussion}). Overall, we infer that $\Lya$ would likely reinforce the emerging picture of (locally) efficient bursts of star formation and rapid outflows in galaxies at cosmic dawn. In follow-up work we will systematically explore similar quantities across a range of star cluster-forming environments, compare more closely to predictions from idealized models of $\Lya$ feedback, and use these to build improved subgrid models to incorporate $\Lya$ feedback in cosmological galaxy formation simulations. Our work demonstrates that $\Lya$ radiation pressure is a critical feedback mechanism to capture on-the-fly in the high-$z$ Universe, which we anticipate will be realized as efficient numerical methods are developed toward this end \citep[e.g.,][]{nebrin_lyman-alpha_2026}

\begin{acknowledgments}
We would like to thank the anonymous referee for a careful reading of the manuscript and their insightful suggestions. SHM would like to thank Lachlan Lancaster, Mike Grudic, Julianne Dalcanton and Shivan Khullar for productive discussions and suggestions. We thank Crystal Martin and Harley Katz for organizing the KITP First Billion Years follow-on workshop where this project was initiated, Avishai Dekel for encouraging this project, and Olof Nebrin for numerous insights and comments on the paper. We acknowledge high-performance computing resources provided by the Simons Foundation as part of the CCA at the Flatiron Institute. The Flatiron Institute is a division of the Simons Foundation. Support for program JWST-AR-08709 was provided by NASA through a grant from the Space Telescope Science Institute, which is operated by the Association of Universities for Research in Astronomy, Inc., under NASA contract NAS 5-03127. This research was supported in part by
grant NSF PHY-2309135 to the Kavli Institute for Theoretical Physics (KITP). This research has made use of NASA’s Astrophysics Data System (ADS) Bibliographic Services.
\end{acknowledgments}

\begin{contribution}
SHM was responsible for leading the research concept, analysis and writing of the manuscript. AS is the lead developer of \textsc{colt}, and contributed to the conceptualization, interpretation, and editing of the manuscript.


\end{contribution}

%

\software{\textsc{colt} \citep{smith_lyman_2015}, \texttt{VETTAM} \citep{Menon_2022}, \texttt{PETSc} \citep{PetscConf,PetscRef}, \texttt{yt} \citep{yt}, \texttt{Matplotlib} \citep{matplotlib}, \texttt{Numpy} \citep{numpy}, \texttt{TurbGen} \citep{Federrath_2022}
          }


\appendix

\section{Approaches for the Eddington ratio}
\label{sec:Appendix_fEdd}
\begin{figure}
  \centering
  \includegraphics[width=0.47\textwidth]{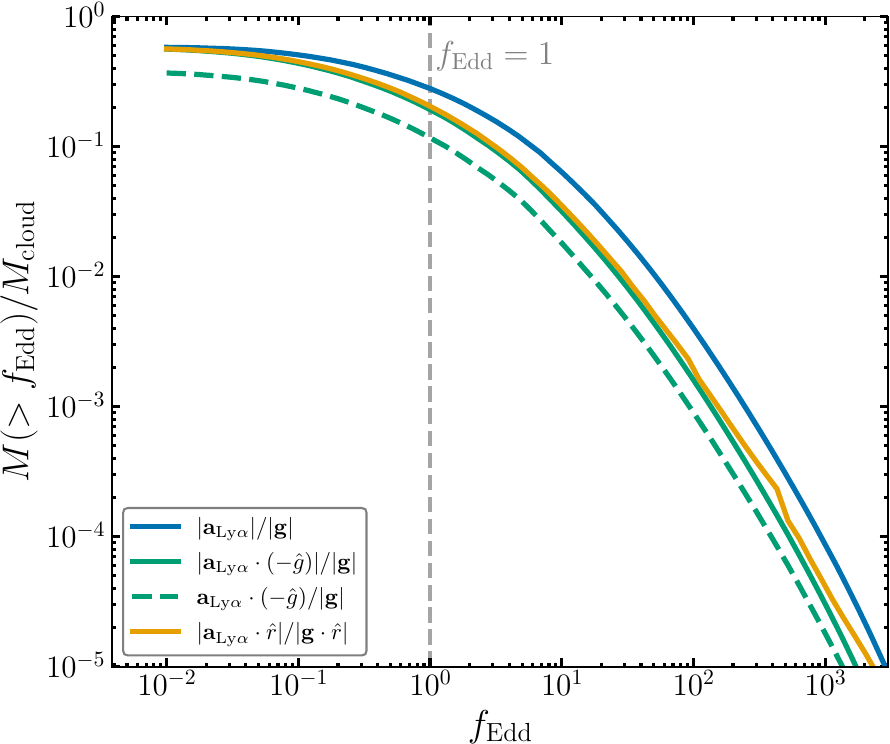}
  \caption{Distributions of the $\Lya$ Eddington ratio, $\fEdd$ (Equation~\ref{eq:fEdd}), shown as the cumulative fraction of $\Mcloud$ with $\fEdd$ greater than a given value at the time $t_\star = 0.83 \, \rm Myr$ for the dust-free case. The four lines correspond to different choices for how $\fEdd$ is calculated, as described in the text of Section~\ref{sec:Appendix_fEdd}. Overall, we find that the different methods agree to within a factor $\sim 2$ to our fiducial choice (green solid) for the super-Eddington mass fraction.}
  \label{fig:Appendix_fEdd}
\end{figure}

\begin{figure*}
  \centering
  \includegraphics[width=0.98\textwidth]{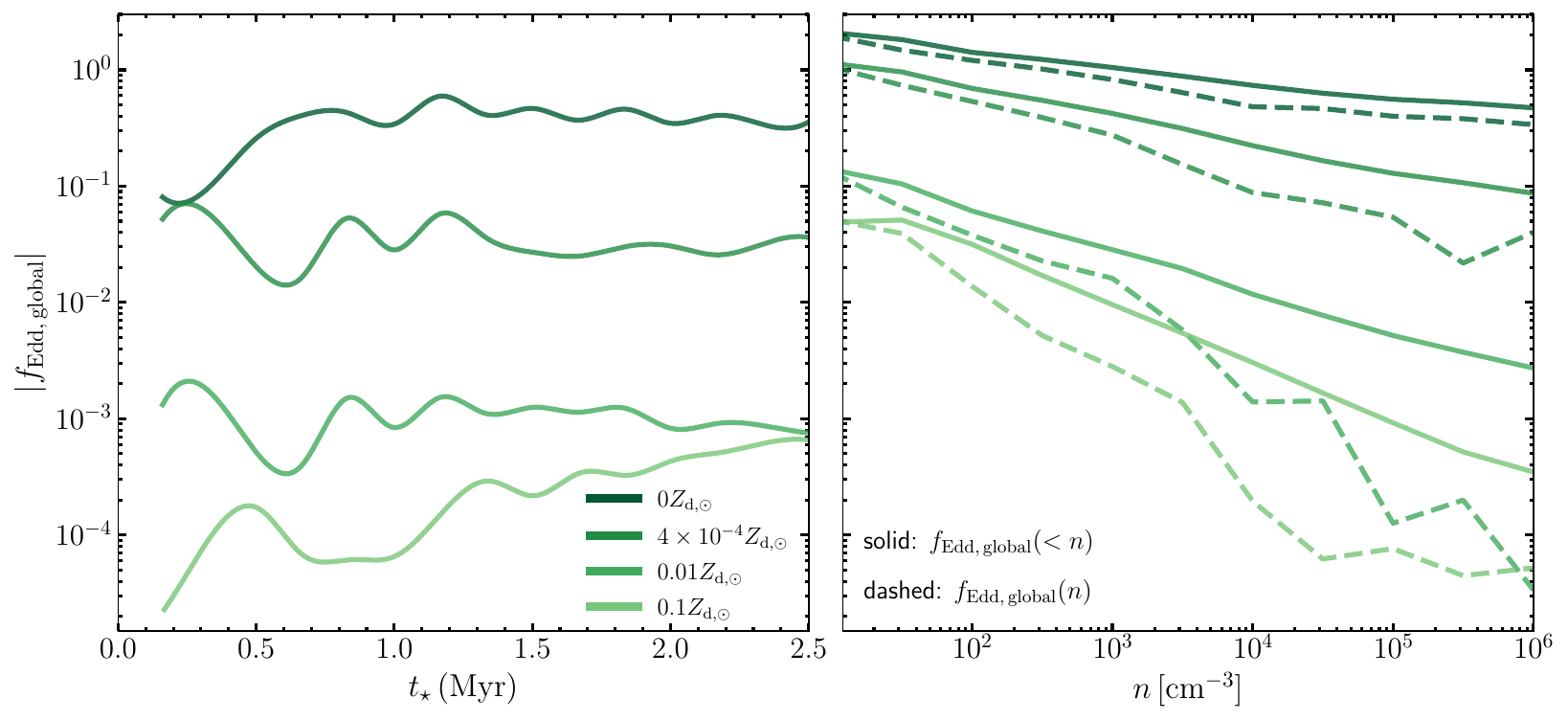}
  \caption{The time evolution of the global Eddington ratio $f_{\rm Edd,global}$ defined as the integrated ratio of the radial $\Lya$ and gravitational forces (Equation~\ref{eq:fEddGlobal}) as a function of stellar age for the four dust cases. We can see that $f_{\rm Edd,global}<1$ consistent with the fact that the bulk of the gas mass is locally sub-Eddington. The influence of the dense gas is reflected in the right panel which shows the result of Equation~\ref{eq:fEddGlobal} when computed over gas with number densities $<n$ (solid) and binned as a function of $n$ (dashed) at the same time as Figure~\ref{fig:fEdd_vs_n}, showing progressively lower $n$ leads to to higher $f_{\rm Edd,global}$.}
  \label{fig:Appendix_fEddGlobal}
\end{figure*}

Here, we quantify the sensitivity of our outcomes to the approach used for computing $\fEdd$. Our fiducial approach is the ratio of magnitudes of the $\Lya$ radiative acceleration projected in the direction of gravity, i.e. Equation~\ref{eq:fEdd}. We also explore three other alternatives: (\textit{i}) the ratio of the magnitude of the accelerations 
\begin{equation}
    \fEdd = \frac{|\aLya|}{|\agrav|} \, ,
\end{equation}
(\textit{ii}) the fiducial case, but considering whether $\Lya$ is oriented parallel or anti-parallel to the gravity
\begin{equation}
    \fEdd = \frac{\aLya \cdot -\agrav}{|\agrav|^2} \, ,
\end{equation}
and (\textit{iii}) projecting both forces in the radial direction, where the radial basis vector ($\hat{\mathbf{r}}$) is calculated with respect to the center of mass/light of the stellar population
\begin{equation}
    \fEdd = \frac{|\aLya \cdot \hat{\mathbf{r}}|}{|\agrav \cdot \hat{\mathbf{r}}|} \, .
\end{equation}
The first alternative has the advantage of simplicity but without carrying any vector information. The second case is identical to our fiducial case, but ignoring super-Eddington gas that has components aligned with the local gravitational acceleration. And the third case is an attempt to describe the radial force competition as would be seen by a single \textit{effective} source, although we note that the distribution of sources is quite distributed. Figure~\ref{fig:Appendix_fEdd} shows that the distribution of $\fEdd$ values are within a factor $\sim 2$ of our fiducial choice, with versions including more vector information having lower $\fEdd$ values, as expected. Overall, this suggests that our takeaways are relatively insensitive to our choice of defintion for the local $\fEdd$.

In addition to this, we also experiment with a global integrated definition of the Eddington ratio $f_{\rm Edd,global}$. In this case, we compute the ratio of \textit{integrated} radial forces of $\Lya$ pressure and gravity as seen by the center of mass of the stellar population, i.e.
\begin{equation}
    \label{eq:fEddGlobal}
    f_{\rm Edd,global} = \frac{\int \rho\,\aLya \cdot  \hat{\mathbf{r}} \, \text{d}V}{\int \rho\,\agrav \cdot  \hat{\mathbf{r}} \, \text{d}V} \, .
\end{equation}
In Figure~\ref{fig:Appendix_fEddGlobal} we show $f_{\rm Edd,global}$ as a function of stellar age ($t_\star$) for the four dust cases. We can see that the global radial force balance is always in favor of gravity for all dust cases---this is reflective of the fact that the bulk of the mass, which would dominate the contribution to $f_{\rm Edd,global}$, is sub-Eddington, and therefore consistent with the local $f_{\rm Edd}$ analysis we follow in this study. In the right panel, we confirm the role of the dense gas in biasing $f_{\rm Edd,global}$ by computing the integrals in Equation~(\ref{eq:fEddGlobal}) cumulatively over gas with number densities $<n$ (solid) and binned as a function of $n$ (dashed). We can see that $f_{\rm Edd,global}$ progressively increases with lower $n$. However, $f_{\rm Edd,global}$ misses the diversity of outcomes of the gravity-feedback competition across the broad density distributions achieved in the gas which can locally regulate collapse and facilitate gas ejection. Due to this reason, we are inclined to adopt locally-defined estimators for $\fEdd$.

\section{Impact of numerical choices}
\label{sec:Appendix_numerical}

\begin{figure}
  \centering
  \includegraphics[width=0.47\textwidth]{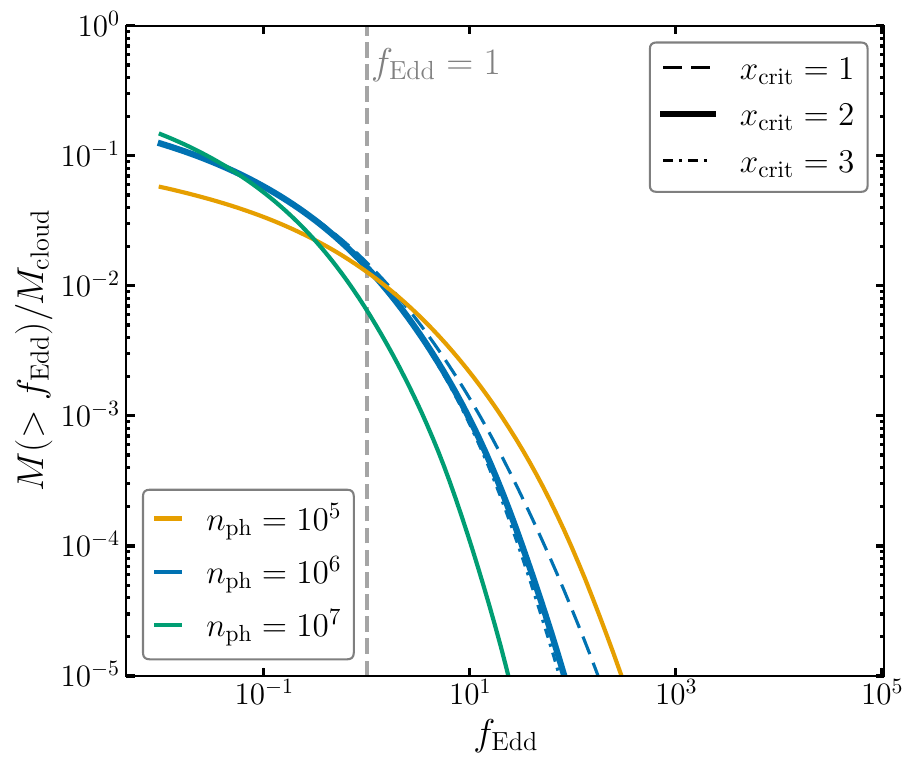}
  \caption{Distributions of the $\Lya$ Eddington ratio, $\fEdd$ (Equation~\ref{eq:fEdd}), shown as the cumulative fraction of $\Mcloud$ with $\fEdd$ greater than a given value at the time $t_\star = 0.83 \, \rm Myr$ for the $Z_{\rm d} = 0.01 \, {\rm Z}_{\rm d,\sun}$ case. Colors show the implications of the number of MC photons ($n_{\rm ph}$) used to sample the radiation field, and linestyles show different choices of $x_{\rm crit}$. We can see that our fiducial case ($n_{\rm ph} = 10^6$, $x_{\rm crit} =2$; blue solid) is converged with respect to $x_{\rm crit}$, but can overestimate $\fEdd$ by a factor few.}
  \label{fig:Appendix_MC}
\end{figure}

In this section we quantify the dependence of $\fEdd$ on the numerical choices we adopt for the MC transport of $\Lya$ photons: namely, the critical core-skipping frequency $x_{\rm crit}$ and the number of MC photon packets ($n_{\rm ph}$). In Figure~\ref{fig:Appendix_MC}, we show the fraction of instantaneous gas mass with $\fEdd$ greater than a given value---similar to Figure~\ref{fig:fEddMainPlot}---at a time $t_\star = 0.8 \, \rm Myr$ compared for for $\xcrit \in \{1,2,3\}$ and $n_{\rm ph} \in \{10^5,10^6,10^7\}$. We can see that the curves are relatively independent of $\xcrit$: while a more accurate choice (i.e. lower $\xcrit$) leads to slightly larger mass fractions with very high $\fEdd$, the total fraction of super-Eddington mass (i.e. $\fEdd>1$) is converged with respect to $\xcrit$. On the other hand, there are noticeable differences in this fraction for different $n_{\rm ph}$, differing by factors of $\lesssim 2$, with lower values of super-Eddington mass at higher values of $n_{\rm ph}$. Therefore more accurate numerical choices (higher $n_{\rm ph}$) only weaken the effect of $\Lya$, reinforcing our conclusions that it is unlikely to alter the gravity-feedback competition for the systems considered in this study.

\section{Semi-analytic star formation model}
\label{sec:Appendix_TK16}

To complement the results of our analysis, we estimate the potential impact of $\Lya$ radiation pressure on the star formation history and final efficiency, using the semi-analytic model framework presented in \citet{Thompson_Krumholz_2016} (hereafter TK16). TK16 models the evolution of an isolated, self-gravitating, turbulent collapsing cloud of mass $M_{\rm cloud}$, considering the competition between momentum injection by stellar feedback and gravity over the broad distribution of gas column densities seen by a point source of radiation representing the (forming) stellar population. This competition is quantified through an Eddington-like approach through the use of the critical surface density $\Sigma_{\rm crit}$ \citep{Thompson_Krumholz_2016}, the maximum column that is super-Eddington for a given momentum injection rate given by,
\begin{equation}
  \Sigma_{\rm crit} = \frac{\langle \dot{p}/M_\star \rangle}{4\pi G},
  \label{eq:Sigma_crit_TK}
\end{equation}
where $\langle \dot{p}/M_\star \rangle$ is the time-averaged specific momentum injection rate of the stellar population, and $G$ is the gravitational constant. This can be used to quantify the mass fraction of super-Eddington sightlines (Eq.~16 of TK16)
\begin{equation}
  \zeta_{\rm m}(x_{\rm crit}) = \frac{1}{2}\left[1 - \mathrm{erf}\!\left(\frac{\sigma_{\ln S}^2 - 2x_{\rm crit}}{2\sqrt{2}\,\sigma_{\ln S}}\right)\right] \, ,
\end{equation}
where $\sigma_{\ln S}$ is the dispersion of the lognormal gas column density distribution (Eqs.~12 and~14 of TK16) to approximate the distribution the stellar source sees over solid angles. $\sigma_{\ln S}$ depends on the turbulent Mach number $\mathcal{M}$ and the slope of the turbulent power spectrum, which we assume to be $k \sim -2$ based on constraints from isothermal turbulence simulations \citep{Federrath_2021}. $x_{\rm crit}$ is the critical log-surface-density threshold
\begin{equation}
  x_{\rm crit} = \ln\!\left(\frac{\Sigma_{\rm crit}}{\Sigma_{\rm gas}}\,\frac{M_\star}{M_\star + M_{\rm g}}\right) \, .
  \label{eq:xcrit_TK}
\end{equation}
Here $\Sigma_{\rm gas} = M_{\rm gas}/(\pi R_{\rm cloud}^2)$\footnote{TK16 consider variations of the model under different assumptions of how the radius of the cloud evolved with time/mass---we choose to keep the radius fixed in our calculations.} is the instantaneous gas mass surface density, $M_\star$ the instantaneous stellar mass, and the factor $M_\star/(M_\star + M_{\rm g})$ accounts for the fact that the radiative momentum is sourced by the stellar mass while gravity is due to both stars and gas. Given this approach, and assuming that gas converts to stars with an efficiency $\epsilon_{\rm ff}$ per free-fall time $t_{\rm ff}$, we obtain the following coupled ODEs for $M_\star$, $M_{\rm gas}$ and the ejected mass through the super-Eddington sightlines $M_{\rm ejected}$

\begin{align}
  \frac{dM_{\rm gas}}{dt} &= -\frac{M_{\rm gas}}{t_{\rm ff}} \left(\epsilon_{\rm ff} + \zeta_{\rm m}\right) \, , \\
  \frac{dM_{\rm *}}{dt} &= \epsilon_{\rm ff} \frac{M_{\rm gas}}{t_{\rm ff}} \, , \\
  \frac{dM_{\rm ejected}}{dt} &= \zeta_{\rm m} \frac{M_{\rm gas}}{t_{\rm ff}} \, .
\end{align}

Thus, the model is described by the parameters $\epsilon_{\rm ff}$ and $\mathcal{M}$, given a model for $\langle \dot{p}/M_\star \rangle$ for the young stellar population. TK16 considers momentum injection due to radiation pressure on dust, in which case $\langle \dot{p}/M_\star \rangle \sim L_{\rm UV}/c$ where $L_{\rm UV}$ is the bolometric UV luminosity coming from the stellar population of mass $M_\star$. We modify this to incorporate the additional momentum injection from $\Lya$ radiation pressure through the definition
\begin{equation}
  \langle \dot{p}/M_\star \rangle = \frac{L_{UV} \left( 1+f_{\rm trap,\Lya} \right)}{M_\star} \, ,
\end{equation}
where $f_{\rm trap,\Lya}$ is our attempt to capture the \textit{boost} over the single-scattering momentum injection from stellar UV photons from resonantly scattered $\Lya$ radiation pressure---similar in principle to the $f_{\rm trap, IR}$ used in the community for the \textit{boost} provided by the dust-reprocessed IR radiation pressure. 

We solve the ODEs in the model using the boundary conditions (i.e.\ at $t=0$) based on the cloud (of $R_{\rm cloud} = 10 \, \rm pc$) modeled in our simulations: i.e.\ $\Mcloud = 10^6 \, \Msun$, and $M_{*} = M_{\rm ejected} = 0$; given the velocity dispersion in the turbulent cloud, and assuming an average gas temperature of $\sim 100 \, \rm K$, we obtain a Mach number of $\mathcal{M} = 20$, although other choices of $\mathcal{M}$ within factors of 2--3 change the outcomes only be a few percent. Our simulations indicate values of $\epsilon_{\rm ff} \sim 30\%$; in addition to this we also compute the model for $\epsilon_{\rm ff} = [10,20] \%$, as our simulations might overestimate $\epsilon_{\rm ff}$ since they lack magnetic fields and assume idealized initial conditions. Nevertheless, since we are primarily interested in the \textit{relative} change in the model outcomes with the inclusion of $\Lya$, our takeaways are insensitive to these model parameter choices. We use the \texttt{BPASS} model outputs for the (age-dependent) value of $L_{\rm UV}$ for the stellar population used in our simulations, and use the values of $f_{\rm trap,\Lya}$ (depending on the dust abundance) obtained from our analysis and shown in the right panel of Figure~\ref{fig:Mf_vs_time}. 

\section{Impact of Monte Carlo noise}
\label{sec:Appendix_MCNoise}

\begin{figure}
  \centering
  \includegraphics[width=0.47\textwidth]{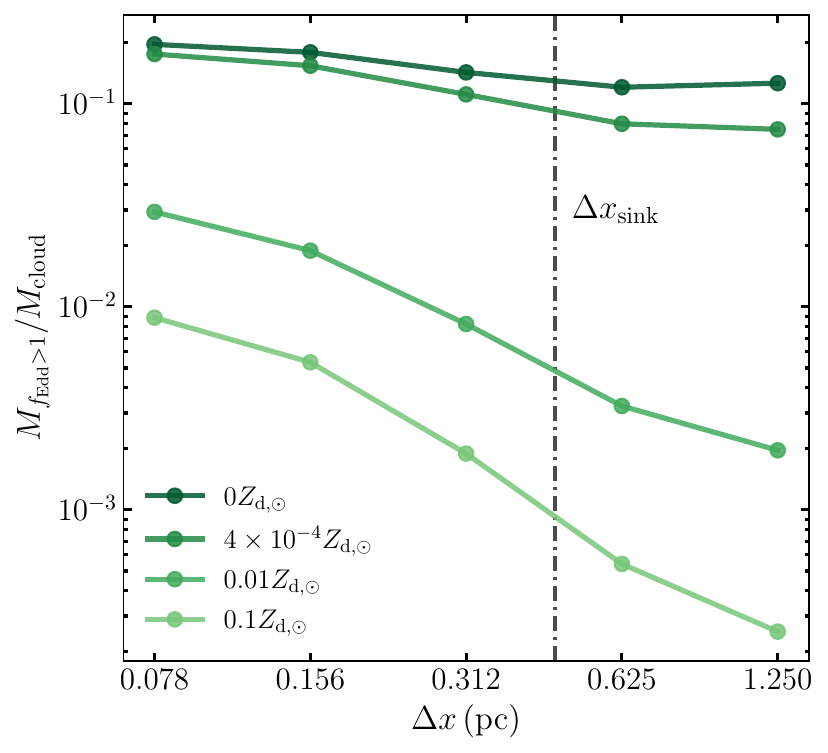}
  \caption{Fraction of cloud mass with $f_{\rm Edd}>1$ (i.e. $M_{\fEdd>1}/M_{\rm cloud}$, as in Figure~\ref{fig:fEddMainPlot}) as a function of grid spacing $\Delta x$, for successively coarsened (block-averaged) versions of the native $512^3$ grid, at the snapshot where the dust-free run's native-resolution mass fraction peaks ($t_\star = 1.16,{\rm Myr}$). Colors follow the dust-treatment convention of Figure~\ref{fig:fEddMainPlot} with the dash-dotted vertical line ($\Delta x_{\rm sink}$) marking the median nearest-neighbor separation between stellar sources, indicating the scales above which a decrease is expected from coarse-averaging due to force cancellation from multiple sources.}
  \label{fig:Appendix_MCNoise}
\end{figure}

In this section we conduct an experiment to quantify the effects of MC noise on our findings for the competition between gravity and $\Lya$ radiation pressure. Specifically, we systematically increase the signal-to-noise of the $\Lya$ radiation field solution and compare the obtained fraction of mass with $\fEdd>1$, i.e., the quantity plotted in Figure~\ref{fig:fEddMainPlot}. To boost the signal-to-noise without increasing the number of photons\footnote{In Figure~\ref{fig:Appendix_MC} we tested for convergence in $\fEdd$ with the number of photons up to $10^7$ photons---an upper limit for what is computationally feasible with resources available to us.}, we progressively coarsen the native $512^3$ computational domain by factors of $\sim 2$ and recompute $\fEdd$ on progressively block-averaged (coarse-grained) versions of the grid. For a coarsening factor $n$, every $n^3$ block of contiguous fine cells is reduced to a single coarse cell with the effective Ly$\alpha$ and gravitational acceleration vectors combined via a mass-weighted average over the $n^3$ fine cells in each block (i.e. preserving momentum conservation). We then reevaluate $\fEdd$ using the fiducial estimator used in our study (Equation~\ref{eq:fEdd}) and compute $M_{\fEdd>1}/M_{\rm cloud}$ for the coarsened versions. We do this for $n=2,4,8,16$, corresponding to effective resolutions of $256^3$, $128^3$, $64^3$, and $32^3$ cells (cell sizes $\Delta x \approx 0.16$–$1.25,{\rm pc}$, compared to $\Delta x \approx 0.08,{\rm pc}$ at native resolution).

In Figure~\ref{fig:Appendix_MCNoise} we compare $M_{\fEdd>1}/M_{\rm cloud}$ for the coarsened versions with the native resolution for all dust cases. We find that $M_{\fEdd>1}$ is relatively insensitive (order unity) to resolution (i.e. signal-to-noise) for low $Z_{\rm d} \lesssim 0.01\,{\rm Z}_{\rm d,\sun}$, especially when considering that a decrease in acceleration is expected at scales larger than the typical stellar source separation ($\Delta x_{\rm sink}$; dash-dot line in Figure~\ref{fig:Appendix_MCNoise}) due to physical cancellation of vectors when averaging (i.e. source confusion). On the other hand, for the dusty cases ($Z_{\rm d} \gtrsim 0.01\,{\rm Z}_{\rm d,\sun}$), where super-Eddington mass fractions are already negligible ($\lesssim 3\%$), decreasing the resolution decreases $M_{\fEdd>1}$ even more (by factors upto $\sim 5$) even at scales $\lesssim \Delta x_{\rm sink}$, suggesting that their dynamical effects are further weakened with higher signal-to-noise spatially-smoothed solutions. Taken together, this experiment reveals that our takeaways are robust---and likely even reinforced---at higher signal-to-noise: $\Lya$ radiation pressure in dust-poor environments can render noticeable fractions of the cloud mass super-Eddington while it is unlikely to affect the star cluster's dynamical competition with gravity in dusty environments. 
%


\bibliography{sample701,lyalpha}{}
\bibliographystyle{aasjournalv7}



\end{document}